%% file: manuscript.tex
\documentclass[superscriptaddress,amsmath,amssymb,aps,pre,nofootinbib]{revtex4-2} 

\usepackage{epsfig,dsfont,amssymb,amsmath,amsthm,amsfonts,amsbsy,mathrsfs} 
\usepackage{graphicx}
\usepackage{color} 
\usepackage{bm}
\usepackage{multirow} 
\usepackage{natbib} 
\usepackage{rotating}
\usepackage{appendix}
\usepackage{mathtools}

\usepackage{tikz}
\usetikzlibrary{arrows}

\newcommand{\beq}{\begin{equation}} 
\newcommand{\eeq}{\end{equation}} 
\newcommand{\bea}{\begin{eqnarray}} 
\newcommand{\eea}{\end{eqnarray}}

\newcommand{\g}{\varphi}

\newcommand{\psiD}{\psi_D}
\newcommand{\psiDk}{\psi}

\newcommand{\MSD}{\langle x^2(t) \rangle}
\newcommand{\MSDLT}{\langle x^2(s) \rangle}

\begin{document}

\title{Unified description of random motions and generalized Fokker-Planck equations}

\author{Luca Angelani}
\email{luca.angelani@cnr.it}
\affiliation{Istituto dei Sistemi Complessi - Consiglio Nazionale delle Ricerche, Piazzale A. Moro 5, I-00185, Roma, Italy
}%
\affiliation{Dipartimento di Fisica,
Sapienza Universit\`a di Roma, Piazzale A. Moro 5, I-00185, Roma, Italy
}%

\begin{abstract}
This review surveys diverse types of random motion within a unified framework, 
introducing essential
mathematical tools, including special functions and fractional calculus, in a pedagogical manner.
We focus on generalized random motions in one-dimensional space described by Fokker-Planck like 
equations, whose fundamental solutions take a universal and simple form 
in the Laplace-Fourier domain in terms of a generic function $\g(s)$. 
By exploring the functional forms of $\g(s)$ 
consistent with physical constraints, we derive a broad class of stochastic dynamics.
This includes Brownian and active motion, simple and anomalous diffusion, and processes featuring stochastic resetting or trapping mechanisms. 
For each case, we provide the corresponding kinetic equation, explicit solutions (where available), and mean-square displacements alongside their asymptotic behaviors. 
Finally, we discuss the physical interpretation of these motions through the lens of time-changed processes and subordinators.
\end{abstract}

\maketitle

\tableofcontents


\section{Introduction}
Random motions are widespread in nature, from the diffusion of molecules
in an aqueous environment or of pollutants through the atmosphere to 
erratic movements of cells and animals \cite{klafter2011first,RWinBio_Berg,Ecology,Hughes_Book_Vol1,Bressloff_book}.
Moreover, random walk models offer a useful tool to describe many phenomena in very different fields of research, from physics
to chemistry, biology, ecology, economics and many others.
There are several approaches to deal with this issue, such as, for example, those based on stochastic Langevin equations and the Ito calculus, L\'evy processes, continuous-time random walks (CTRW), 
Fokker-Planck equations \cite{klafter2011first,Gardiner,applebaum2009levy,Risken,NSM_Book2025}.
The latter approach offers a very general and powerful framework for dealing 
with random motions in the most diverse conditions and situations, e.g., in non-homogeneous cases, in the presence of external forces, in confining domains with different  boundary conditions. 
Interest in this longstanding, fundamental topic continues to grow, driven by recent developments in highly promising areas with broad scope and wide-ranging applications.
Here are some of them.

{\it Active matter} describes systems composed of elements which locally convert environmental energy into
systematic motion, thus resulting in  intrinsically out-of equilibrium phenomena.
The past two decades have seen a tremendous increase in publications on the active matter subject, 
covering a wide range of disciplines -- see, e.g., 
the recent overview of the existing review articles
on this topic \cite{VW_AM_2024}.
The constitutive elements of active systems can be living organisms, such as cell, bacteria and animals, or non-living
objects, such as self-catalytic colloids, Janus particles, vibrating granular materials and micro-robots 
\cite{Marchetti_2013,Elgeti_2015,Play_AM_2024}.
Many interesting phenomena emerge in such a kind of systems, from collective behaviors at large scales
to unusual non-equilibrium effects at the single particle level, such as accumulation at boundaries 
or rectification of motion -- 
see, e.g., Ref. \cite{RevModPhys.88.045006} and references therein.
Persistent random walks are the typical motions of active particles. 
For example, motile flagellated bacteria, such as {\it E.coli},  
swim using a run-and-tumble strategy, alternating a run phase at constant speed along straight lines to
a tumble phase in which the cell reorients its direction of motion
\cite{Ecoli_Berg,RWinBio_Berg,PhysRevE.48.2553}.
This type of random walk, 
which in one dimensional space is related to the telegraph equation \cite{weiss2002some},
possesses many interesting features, and its study in a wide variety of situations 
has led to important advances in the understanding of non-equilibrium phenomena in active matter
\cite{Cates_2012,PhysRevLett.100.218103,MIPS_2015,Angelani2024EPJE,angelani2023one,Fodor_2016}, such as , for example, in the
recent studies on time-irreversibility and entropy production 
\cite{Time_Irrev_2022,PhysRevE.105.034113,Cocconi2020,GarciaMillan_2021,Razin2020,Angelani_Entropy2024,EPR_PRE2025}.

{\it Anomalous diffusion} refers to random motions that exhibit, at long time, 
power-law behavior of the mean-square displacement (MSD), $\MSD \propto t^\mu$, with generic exponent $\mu>0$.
Normal diffusion corresponds to $\mu=1$, ballistic motion to $\mu=2$, while non integer values of $\mu$ 
describe anomalous behaviors:
subdiffusive if $\mu<1$ and superdiffusive if $\mu>1$.
Anomalous diffusion is present in a broad class of phenomena \cite{Klafter_2005,Levy_walks_2015,Vilk_2022,Sandev_Chaos_2026},
from transport processes through amorphous solids and disordered media \cite{scher1975anomalous,DDM}
to animal dispersion \cite{ForagingBook,Ecology} and financial stock price variations \cite{PhysRevE.62.R3023}.
A useful and powerful tool for dealing with anomalous diffusion is based on fractional diffusion equations and,
more generally, fractional Fokker-Planck equations 
\cite{metzler2000random,klafter2011first,C4CP03465A,Evangelista_Lenzi_2018,SoKl_2005,JM_2016}.
These are integro-differential equations nonlocal in time (or in space, not considered here) 
in which there are power-law kernels 
describing slowly decaying memory processes.
Fractional calculus has proved to be of great importance in dealing with anomalous diffusion processes 
\cite{metzler2000random,gorenflo2020mittag,kilbas2006theory}.

{\it Stochastic resetting} describes stochastic processes in which the system,
during its evolution, has a finite probability of resetting to its initial condition.
It is typically associated with search processes in which one must account for the strategy of returning to the starting point following an unsuccessful search.
It has, however, a wide range of applications -- see, e.g., Ref.s \cite{Evans_2020,Gupta_2022,Montero_2017}
and references therein.
A key property of resetting processes is the existence of a non-equilibrium stationary state. 
The problem of resetting for a diffusive particle was studied by Evans and Majumdar 
\cite{Evans_DSR,Evans_2011}
and after that many works addressed the problem in  a variety of different systems,
such as, e.g., telegraphic and run-and-tumble models \cite{Evans_2018,Masoliver_2019,Bress_2020,Tucci_2022,Santra_2020},
anomalous diffusion \cite{KGN_2019,Maso_ADSR,Singh_2022,Shkilev_2022},
general 
CTRW \cite{Pagnini2026},
fractional telegraph equation \cite{Gor_2024,Sandev_2024},
comblike structures \cite{Doma_2020}, just to mention a few recent investigations.

{\it Trapping} processes occur when motions take place in complex environments that reduce
or inhibit the particle's ability to move. Trapping can be transient or irreversible.
The motion of colloidal particles or cells in the presence of obstacles, through porous media, under external confining
potentials or subject to geometrical constraints provides typical examples of the former;
molecules absorption by reactive substrates, cell death in harmful environments, 
irreversible bacterial adhesion to surfaces during biofilm formation are examples of the latter 
-- see, e.g., Ref.s \cite{PhysRevLett.110.220603,Angelani_PS2023,KOS_PRE2023,Doerries_2023,peruani_2025} and references therein.
Modeling random motions in trapping environments is therefore of great relevance.
Irreversible trapping has been recently investigated in different contexts,
considering diffusive motions in the presence of a sequence of absorbing traps
\cite{Pozzoli_2021,Pozzoli_2022}, subdiffusion with particle immobilization \cite{KOS_PRE2023},
run-and-tumble motions through trapping regions of various extension \cite{Angelani_PS2023}.

\medskip
In this work I treat in a unified way a class of random motions in one-dimensional space 
described by a generic integro-differential 
equation whose fundamental solution takes a very simple form, 
i.e., a
Lorentzian shape of the Fourier-Laplace transformed 
probability density function (PDF)
as a function of $k$, see equation (\ref{Pks}) below.
This depends on a generic function $\g(s)$ encoding the property of the random motion.
In the  spirit of previous studies on generalized diffusion and telegrapher's equations 
\cite{Sandev2017,SMC2018,GBM2020,Gorska_PRE2020,Gorska_2023,Tomovski_2025}, 
I will systematically examine all possible choices of the function $\g$ that satisfy the relevant physical constraints, including the non-negativity of the PDF, the prescribed initial conditions, and the existence or absence of a non-trivial stationary solution. 
I will then show how this framework allows us to describe a broad range of stochastic motions, from simple diffusion to run-and-tumble dynamics, stochastic resetting, trapping processes, and anomalous diffusion.
For each of them the corresponding Fokker-Planck like equation is reported, together with the 
expression of the MSD and its asymptotic behavior. 
The present approach allows us to reproduce well known results in the literature, extends previous
findings to a broader class of systems and obtain in some cases new formalizations and results 
of topics that are on the edge of current research.
Furthermore, it provides a simple and pedagogical overview of  Fokker-Planck-like equations for different types of random motion within a unified framework, complemented by a compendium of essential mathematical tools, such as fractional calculus and Bernstein functions.

\medskip
The paper is organized as follows. \\
In Section \ref{Sec2} we introduce the class of random motion 
that we analyze in this work. 
We report the generalized Fokker–Planck–like equation and the corresponding fundamental solution, discussing the conditions under which it represents a physically meaningful probability density function of a random process. We also provide the expression for the mean squared displacement and analyse its asymptotic behaviour.\\
In Sections \ref{Sec3} we treat in details the different types of random motions: 
diffusion (\ref{Sec_D}),
ballistic (\ref{Sec_B}),
diffusion with resetting or trapping (\ref{Sec_DR}),
anomalous diffusion (\ref{Sec_AD}),
anomalous diffusion with resetting (\ref{Sec_ADR}),
anomalous diffusion with trapping (\ref{Sec_ADT}),
run-and-tumble (\ref{Sec_RT}),
run-and-tumble with resetting or trapping (\ref{Sec_RTR}),
anomalous run-and-tumble (\ref{Sec_ART}),
anomalous run-and-tumble with resetting (\ref{Sec_ARTR}),
anomalous run-and-tumble with trapping (\ref{Sec_ARTT}).\\
In Section \ref{Sec_Sub} we give a brief overview of how the various random motions treated before can be interpreted in terms of time-changed parent processes in the framework of L\`evy processes and subordinators.\\
The conclusions are presented in Section \ref{Sec_Concl}, together with a discussion of possible extensions and perspectives. \\
We include some useful mathematical background in the Appendix, introducing and defining some special functions, Fourier and Laplace transforms, completely monotone and Bernstein functions and a very short compendium of fractional calculus. \\

\section{Generalized random motions}
\label{Sec2}
We consider a generic one-dimensional random motion starting at the origin. 
The single-time PDF,  
$P(x,t)$, of finding the particle at position $x$ at a given time $t$
satisfies the generalized Fokker-Planck equation 
\begin{equation}
\label{FPeq}
T P(x,t) = v^2 \partial_x^2 P(x,t) ,
\end{equation}
where $T$ is a generic linear integro-differential operator acting on the time variable $t$
that defines the specific class of random motion.
For example, $T$ can reduce to standard local derivatives 
such as  $\partial_t$ (Brownian motion), $\partial_t^2$  (ballistic motion) or 
their linear combination $\partial_t^2+\alpha \partial_t$ (telegraph equation, run-and-tumble models).
Alternatively, $T$ can manifest as a non-local time-integral operator describing transport processes with memory (anomalous diffusion models), which often results in a fractional derivative operator 
$\partial_t^\mu$ 
(see below)\footnote[1]{We denote with $\partial_t^n$ the $n$-th partial derivative operator 
$\partial^n /\partial t^n$ and with
$\partial_t^\mu$ the partial Caputo fractional derivative of non-integer order $\mu$ (see Appendix).}.
We stress that our analysis concerns drift-free stochastic processes in the absence of external potentials. 
Without loss of generality, $v^2$ denotes a non-negative quantity 
acting as a diffusivity constant, the square of the particle's speed, or a generic parameter with appropriate physical dimensions depending on the model.
We note that while a single-time PDF does not completely characterize a generic non-Markovian process (which fundamentally requires multi-time joint distributions to resolve its history) the generalized framework of Eq. (\ref{FPeq}) encompasses both Markovian and non-Markovian dynamics. In the latter case, the historical dependencies of the process are effectively encoded into the memory kernel of the non-local operator $T$, governing the exact evolution of the marginal distribution 
\cite{klafter2011first,Sokolov2002}. \\
We study fundamental solutions of (\ref{FPeq}), i.e., considering initial conditions
\begin{equation}
\label{ic}
    P(x,t)|_{t=0} = \delta(x) ,
\end{equation}
and, depending on the nature of the operator $T$ defining the random motion,
also 
\begin{equation}
\label{ic2}
\partial_t P(x,t)|_{t=0}=0 .
\end{equation}
We consider here operators $T$ that, when applied to a function $P$, have a Laplace-Fourier transform that can be written as
\begin{equation}
\label{Top}
    \left( {\cal F} \circ {\cal L}\right)  [T P] (k,s) = \g(s) \hat {\tilde P}(k,s) +\chi(s) 
\end{equation}
where ${\cal F}$ and ${\cal L}$ (and symbols $\hat{P}$ and $\tilde{P}) $ denote, respectively, the Fourier and Laplace transforms (see Appendix), and
$\g(s)$ and $\chi(s)$ are generic functions.
As will be shown, this choice enables the description of most physically meaningful processes, by ensuring causality and incorporating time-invariant kernels, such as in  the generic form 
$T[f](t) = \int_0^t dt' K(t-t') f^{(n)}(t')$, with $f^{(n)}$ the $n$-th derivative of $f$.
Therefore, the generalized Fokker-Planck equation (\ref{FPeq}), in the Laplace-Fourier domain, reads
$$\g(s) \hat {\tilde P}(k,s) +\chi(s)= - v^2 k^2 \hat{\tilde P}(k,s). $$
By requiring the PDF $P$ to be normalized, i.e., $\hat{\tilde P}(k=0,s)=1/s$ (see equation (\ref{NCE}) below),
we have that $\chi(s)=\g(s)/s$ and, then, operators $T$ can be written in terms of a single generic function $\g(s)$ as follows
\begin{equation}
\label{FLT}
    \left( {\cal F} \circ {\cal L}\right)  [T P] (k,s) = 
    \frac{\g(s)}{s} \left( s{\hat {\tilde P}}(k,s)- 1     \right) .
\end{equation}
We note that, from (\ref{FLT}), we can formally write (\ref{FPeq}) in an alternative form
\begin{equation}
    \partial_t P(x,t) = L P(x,t) ,
\end{equation}
where the Laplace-Fourier transform of the linear operator $L$, acting on both  variables $t$ and $x$ (through the operator $\partial_x^2$), reads
\begin{equation}
    \left( {\cal F} \circ {\cal L}\right)  [L P] (k,s) = -\frac{v^2 k^2 s}{\g(s)} 
    {\hat {\tilde P}}(k,s) ,
\end{equation}
and we have considered initial conditions and used the property (\ref{LT_der}).\\
The quantities $v^2$ and $\g(s)$ have physical dimensions such that $[v^2]/[\g(s)]$ has the dimension of length square.
In conclusions, the generalized Fokker-Planck equation (\ref{FPeq}) with initial conditions (\ref{ic})-(\ref{ic2}) reads, in the Laplace-Fourier domain 
\begin{equation}
\label{FPeqFL}
\g(s) \left( s \hat{\tilde P}(k,s) -1\right) = - s v^2 k^2 \hat{\tilde{P}}(k,s) ,
\end{equation}
whose solution is
\begin{equation}
\label{Pks}
    {\hat {\tilde P}}(k,s) = \frac{1}{s} \ \frac{\g(s)}{\g(s)+v^2 k^2} .
\end{equation}
The nature of the random motion is dictated by the form of the function $\g(s)$.
A similar generalized form has been investigated in the context of telegrapher's equation with memory kernels \cite{Gorska_PRE2020,Tomovski_2025}.
It is worth noting that the general expression (\ref{Pks}) can also be formally 
derived in the context of CTRW at large scales, 
see, e.g., chapters 3 and 5 of \cite{klafter2011first}, where the PDF for waiting times  $\psi(t)$ and the integral kernel $M(t)$ presented there are related to our function $\g$ through $\tilde{\psi}(s)=[1+\bar{\g}(s)]^{-1}$ and $\tilde{M}(s)=1/\bar{\g}(s)$, with $\bar{\g}(s)=(\ell^2/v^2)\g(s)$ adimensional quantity and $\ell$ sets the unit  length, $\lambda(k)\simeq 1-\ell^2 k^2$, 
where $\lambda$ is the characteristic function of the displacement per step in the CTRW framework
\cite{klafter2011first}.
As will become clear in what follows, this very general expression will allow us 
to describe most of the known types of simple random motions. 
However, we  note that the above expression does not cover all  possible cases. More complex types of motion require more involved (model specific) 
expressions of PDF. 
For example, just to cite a few, this is the case of  
run-and-tumble motions with finite tumbling times 
-- see, for example, equation (61) in Ref. \cite{Angelani_Ortho} or equation (17) in Ref. \cite{angelani2013averaged} --
or of general intermittent motions 
\cite{angelani2013averaged,Gupta2024}
or the {\it soft} resetting process of diffusion with memory \cite{KGN_2019,Shkilev2017}.\\
The inverse Fourier transform of (\ref{Pks}) gives the PDF in the $(x,s)$ domain
-- see (\ref{FT2}) -- 
\begin{equation}
\label{Pxs}
    {\tilde P}(x,s) = \frac{\sqrt{\g(s)}}{2vs} \ \exp \left( -\frac{|x|}{v} \sqrt{\g(s)} \right) .
\end{equation}
First of all, we note that the PDF verifies the normalization condition. Indeed, from (\ref{Pks}) we have
\begin{equation}
\int_{-\infty}^{+\infty} dx \ P(x,t) = {\hat P}(k=0,t) = 
{\cal L}^{-1} [{\hat {\tilde P}}(k=0,s)](t) = 
{\cal L}^{-1} [1/s](t) = 1 .
\label{NCE}
\end{equation}
As said before, the specific form of the function $\g(s)$ defines the different types of motion. 
However, there are some conditions that the function $\g$ must satisfy to describe random motions.\\
\begin{itemize}
    \item
First, for the solution $P(x,t)$ to be a PDF, it must be non-negative defined, i.e., $P(x,t)\geqslant 0$.
By using the Bernstein’s theorem  
which states that a function is completely monotone 
if and only if it is the Laplace transform of a non-negative measure 
(see Appendix and Ref. \cite{schilling-bern}),
we have that the function ${\tilde P}(x,s)$ defined in (\ref{Pxs}) must be a completely monotone function. 
This determines a condition on the function $\g(s)$, which must be such that 
$\sqrt{\g(s)}$ is a Bernstein function ($\mathcal{BF}$) -- see  properties ({\it ii}), ({\it v}) and ({\it vi}) for completely monotone and Bernstein functions in the Appendix -- i.e. we must require that (Condition 1)
\begin{equation}
\label{c1}
    \sqrt{\g(s)} \in \mathcal{BF} \qquad \qquad {\text{(C1)}}
\end{equation}
\item
Second, the initial condition (\ref{ic}) in the Fourier space reads ${\hat P}(k,0)=1$ and, 
from the initial value theorem (see Tauberian theorem in the Appendix, with $\nu=1$), we must require that
$$
\lim_{t\to 0} {\hat P}(k,t) = 
\lim_{s \to \infty} s {\hat {\tilde P}}(k,s) = 1,
$$
which is satisfied if the function $\g(s)$ diverges when $s \to \infty$ (Condition 2)
\begin{equation}
\label{c2}
    \lim_{s \to \infty} \g(s) = +\infty  \qquad \qquad {\text{(C2)}}
\end{equation}
\item
There is an additional condition on $\g$ that must be considered if we require the PDF $P(x,t)$ to asymptotically spread over the entire space.
By considering solutions for which $P(x,t \to\infty)=0$ for all $x$, i.e. 
$$
\lim_{t \to \infty} {\hat P}(k,t) = \lim_{s \to 0} s{\hat {\tilde P}}(k,s) = 0 ,
$$
this results, from (\ref{Pks}), in a condition for $\g$ (Condition 3a)
\begin{equation}
\label{c3}
    \lim_{s \to 0} \g(s) = 0  \qquad \qquad {\text{(C3a)}}
\end{equation}
However, there are situations in which this is not the case. For example, when considering stochastic resetting or trapping processes (see below), we have non-trivial stationary solutions, i.e.,  
asymptotically $P(x,t \to\infty)>0$,
\begin{equation}
\label{Pst}
    P_{st.}(x) = \lim_{t \to \infty} P(x,t) = 
    \lim_{s \to 0} s {\tilde P}(x,s) = \frac{\sqrt{\g(0)}}{2v} \ \exp \left( -\frac{|x|}{v} \sqrt{\g(0)} \right)
\end{equation}
implying that, in this case, we must require that (Condition 3b)
\begin{equation}
\label{c3b}
    \lim_{s \to 0} \g(s) > 0  \qquad \qquad {\text{(C3b)}}
\end{equation}
We note that Eq. (\ref{Pst}) provides the non-trivial stationary distribution for all the models considered in the general case.
\end{itemize}
Summarizing, the function $\g(s)$ entering in the expression of the PDF, (\ref{Pks}) and (\ref{Pxs}), and defining the type of random motion, 
must satisfy conditions C1 (\ref{c1}), C2 (\ref{c2}) and C3a (\ref{c3}) or C3b (\ref{c3b}) (see Table I).

\medskip
A useful quantity that characterizes the behavior of the system is the mean-square displacement.
It can be calculated from the knowledge of the PDF, through the relation
(see, e.g., Eq. (1.8), in connection with Eq. (1.4), of Ref. \cite{klafter2011first})
\begin{equation} 
\MSD 
= \int_{\mathbb R} dx \ x^2 P(x,t)=- \left. \partial^2_k {\hat P}(k,t) \right|_{k=0} .
\end{equation}
Using the general expression (\ref{Pks}) we obtain, in the Laplace domain,
\begin{equation}
    \label{r2s}
    \MSDLT  = \frac{2v^2}{s \g(s)} .
\end{equation}
This is a simple and very general expression which allows us to determine the analytic form of the MSD
whenever we are able to determine the inverse Laplace transform of $1/s\g(s)$. 
The behavior of the function $\g(s)$ at small and large $s$ determines the nature of the random motion at
long and short times, respectively.
Indeed, from the Tauberian theorem \cite{feller1971,klafter2011first} (see Appendix (\ref{Sub_FLT})),
the asymptotic behavior of the MSD for $t\to \infty$ and $t\to 0 $ can be obtained from the corresponding behavior of its Laplace transform for $s\to 0$ and $s\to \infty$.
For example, if $\g(s) \simeq s^\nu$ (for small or large $s$) we have that 
$\MSDLT \simeq 2v^2 s^{-(1+\nu)}$ and, therefore, the behavior of the MSD 
(at long or short times) is
\begin{equation}
    \MSD \simeq \frac{2v^2}{\Gamma(1+\nu)} \ t^\nu .
\end{equation}
Depending on the value of the exponent $\nu$ we have then diffusive behavior ($\nu=1$), ballistic ($\nu=2$) and anomalous diffusion for non-integer values of $\nu$, namely subdiffusion ($\nu<1$) or superdiffusion ($\nu>1$).

\begin{table}
\centering
\setlength{\tabcolsep}{5pt}
\renewcommand{\arraystretch}{1.5}
 \begin{tabular}{||c| c | c |c | c | c ||} 
 \hline 
 \rule{0pt}{20pt} 
 ${\tilde {\hat P}}(k,s)$ &  ${\tilde P}(x,s)$ & $\MSDLT$ & \multicolumn{3}{|c|}{Conditions on $\g(s)$}  \\ [2ex] 
 \hline 
 \hline 
  \rule{0pt}{23pt} 
$\displaystyle \frac{1}{s} \frac{\g(s)}{\g(s)+v^2 k^2}$ &
$\displaystyle \frac{\sqrt{\g(s)}}{2v s} \exp \left( -\frac{|x|}{v} \sqrt{\g(s)} \right)$ &
$\displaystyle \frac{2 v^2}{s \g(s)}$ &
 $ \sqrt{\g(s)} \in \mathcal{BF}$ &
 $\lim\limits_{s \to +\infty} \g(s) = +\infty $ &
\begin{tabular}{c} 
 $\lim\limits_{s \to 0} \g(s) = 0 $\\ 
 $\lim\limits_{s \to 0} \g(s) > 0 $  
 \end{tabular} \\ [4ex]
\hline
 \end{tabular}
 \caption{General expressions of the PDFs in the $(k,s)$ and $(x,s)$ domains, and of the MSD in the Laplace domain.
 We also report the conditions that the function $\g(s)$ must meet to describe a random motion process with the considered
 initial and asymptotic conditions. See Appendix for the definition of Bernstein functions ($\mathcal{BF}$).}
\end{table}

\section{Types of random motions}
\label{Sec3}

We investigate the different kinds of random motions described by the PDF expression
(\ref{Pks}).
We will proceed following a kind of reverse reasoning, starting from the various simple $\g(s)$ functions 
satisfying physical conditions C1 (\ref{c1}), C2 (\ref{c2}) and C3a (\ref{c3}) or C3b (\ref{c3b}) (see Table I)
and showing how they correspond to Fokker-Planck like equations describing random motions of various kinds: 
from diffusive motions, to run-and-tumble walks, anomalous diffusion, 
motions in the presence of stochastic resetting and irreversible trapping.
The results obtained are summarized in Table II. 

\subsection{Diffusive motion}
\label{Sec_D}
Let us start with the simplest $\g$ function 
\begin{equation}
\boxed{ \g(s) = s }
\end{equation}
It meets condition  (\ref{c1}), as we have that 
$\sqrt{\g(s)}=\sqrt{s} \in \mathcal{BF}$ (see Appendix), and also conditions (\ref{c2}) and (\ref{c3}).
The expression of the PDF in the Laplace-Fourier domain (\ref{Pks}) reads in this case
\begin{equation}
\label{PDF_ks_BM}
 {\hat {\tilde P}}(k,s) = \frac{1}{s+v^2 k^2} ,
\end{equation}
that can be rewritten as 
$$
(s+v^2 k^2) {\hat {\tilde P}}(k,s) =1 .
$$
By performing inverse Fourier transform (see (\ref{FT_der}) with $n=2$ and (\ref{FT1})) we obtain
$$
(s- v^2 \partial_x^2) {\tilde P}(x,s) = \delta(x),
$$
and we finally arrive, by using  inverse Laplace transform,
to the diffusion equation 
(with diffusion constant $D=v^2$)
\begin{equation}
 \partial_t P(x,t) = D \partial_{x}^2 P(x,t) ,
\end{equation}
having used the property of the Laplace transform 
${\cal L}[\partial_t f(t)] = s {\tilde f}(s) - f(0)$ (see (\ref{LT_der}) with $n=1$)
and initial condition (\ref{ic}).\\
The PDF in the $(x,s)$ domain is given by (\ref{Pxs})
\begin{equation}
    {\tilde P}(x,s) = \frac{1}{2\sqrt{Ds}} \exp\left(-|x| \sqrt{\frac{s}{D}}\right) , 
\end{equation}
and the PDF in the $(k,t)$ domain is obtained by using (\ref{PDF_ks_BM}) and the property (\ref{LT_tn2}) with 
$n=0$
\begin{equation}
    {\hat P}(k,t) = \exp \left( -D k^2 t \right) .
\end{equation}
From the previous expression, by using the property (\ref{FT3}), we finally arrive at the Gaussian distribution PDF 
\begin{equation}
\label{Pxt_D}
    P(x,t) = \frac{1}{\sqrt{4\pi D t}} \exp\left(-\frac{x^2}{4Dt}\right) .
\end{equation}

\medskip
\noindent The MSD in the Laplace domain (\ref{r2s}) reads
\begin{equation}
    \MSDLT = \frac{2D}{s^2}  ,
\end{equation}
corresponding, in the time domain, to the diffusive behavior (see (\ref{LT_tn1}))
\begin{equation}
\MSD = 2 D t 
\end{equation}

\subsection{Ballistic motion (wave equation)}
\label{Sec_B}
The second simplest function $\g$ which obeys (\ref{c1}), (\ref{c2}) and (\ref{c3})
is 
\begin{equation}
 \boxed{ 
 \g(s) = s^2
 }
 \end{equation}
Indeed, we have that $\sqrt{\g(s)}=s \in \mathcal{BF}$.
The expression (\ref{Pks}) now reads
\begin{equation}
 {\hat {\tilde P}}(k,s) = \frac{s}{s^2+v^2 k^2} ,
\end{equation}
which can be written as
$$
(s^2+v^2 k^2) {\hat {\tilde P}}(k,s) =s .
$$
The inverse Fourier transform gives
$$
(s^2- v^2 \partial_x^2) {\tilde P}(x,s) = s \delta(x),
$$
and, by noting that ${\cal L}[\partial^2_t f(t)] = s^2 {\tilde f}(s) - s f(0) - \partial_t f(t)|_0 $ 
(see (\ref{LT_der}) in Appendix)
we obtain the wave equation, describing the ballistic motion of a particle 
with speed $v$ satisfying initial conditions (\ref{ic}) and (\ref{ic2})
\begin{equation}
\partial^2_t P(x,t) = v^2 \partial_{x}^2 P(x,t) .
\end{equation}
The solution (\ref{Pxs}) reads
\begin{equation}
    {\tilde P}(x,s) = \frac{1}{2v} \ \exp\left( -\frac{|x|}{v} s \right) ,
\end{equation}
which, in the time domain, is (see (\ref{LT_delta2}))
\begin{equation}
\label{Pxt_B}
    P(x,t) = \frac{1}{2v} \delta\left(t- \frac{|x|}{v} \right) 
    = \frac{\delta(x-vt)+\delta(x+vt)}{2} ,
\end{equation}
that is a free propagation at constant speed $v$ of two delta peaks in opposite directions.\\

\medskip
\noindent
For the MSD we have
\begin{equation}
    \MSDLT = \frac{2v^2}{s^3}  ,
\end{equation}
which, in the time domain, gives a ballistic behavior (see (\ref{LT_tn1}))
\begin{equation}
    \MSD = v^2 t^2 .
\end{equation}

\subsection{Diffusion with resetting or trapping}
\label{Sec_DR}
Let us go further and  consider a  function $\g$ of the form
\begin{equation}
\label{g_DR}
 \boxed{   
 \g(s) = s + r } 
\end{equation}
with $r>0$ a positive parameter.
Also in this case we have that $\sqrt{\g(s)}=\sqrt{s+r} \in \mathcal{BF}$,
as both $\sqrt{s}$ and $s+r$ are $\mathcal{BF}$ and considering the composition property of Bernstein functions
-- see ({\it iii}) in the Appendix.
In this case  (\ref{Pks}) reads
\begin{equation}
\label{Pks_DR}
    {\hat {\tilde P}}(k,s) = \frac{1}{s} \frac{s+r}{s+r+v^2 k^2}, 
\end{equation}
which can be rewritten as 
$$
(s+r+v^2k^2)  {\hat {\tilde P}}(k,s) = 1 +r/s ,
$$
or
$$
[s  {\hat {\tilde P}}(k,s) -1 ] + r [ {\hat {\tilde P}}(k,s) -1/s] = -v^2 k^2  {\hat {\tilde P}}(k,s) .
$$
By performing the inverse Fourier transform we obtain
$$
s {\tilde P}(x,s) - \delta(x) + r [{\tilde P}(x,s) - \delta(x)/s] = v^2 \partial_x^2 {\tilde P}(x,s) ,
$$
and, applying the inverse Laplace transform (see (\ref{LT_der}), (\ref{LT_1sus})), 
we finally arrive at
\begin{equation}
\label{D_R}
\partial_t P(x,t)  = v^2 \partial_x^2 P(x,t) - r P(x,t)  + r \delta(x) .
\end{equation}
{Stochastic resetting.} 
The above equation describes a diffusion process 
in the presence of stochastic resetting \cite{Evans_DSR,SR_Gupta2022}: 
a particle undergoes free diffusion in the entire space and resets its position to the origin at rate $r$.
The last two terms in (\ref{D_R}) correspond, respectively, to the loss of probability at location $x$ and the gain
of it at the origin due to resetting events.
That this describes a resetting mechanism is also evident from the expression (\ref{Pks_DR}), that can be written as
\begin{equation}
\label{BM_ren0}
 {\hat {\tilde P}}(k,s) = \left( 1+ \frac{r}{s} \right) {\hat {\tilde P}}_0(k,s+r) \ , 
\end{equation}
with $P_0$ the PDF of the  diffusion process without resetting (\ref{Pxt_D}).
In the $(x,t)$ domain this leads to the so called renewal equation defining  
resetting processes \cite{Evans_2018}
\begin{equation}
\label{BM_ren}
    P(x,t) = e^{-rt} P_0(x,t) + r \int_0^t d\tau \ e^{-r\tau} P_0(x,\tau) \ ,
\end{equation}
where we have used (\ref{LT_trasl}) and (\ref{LT_conv}),
together with (\ref{LT_1sus}), which expresses the inverse Laplace transform of $1/s$.
Indeed, for the stochastic resetting process we can write the following relation
$$
P(x,t) = e^{-rt} P_0(x,t) + r \int_0^t d\tau \ e^{-r\tau} P_0(x,\tau) \int dx' P(x',t-\tau) \ ,
$$
where the first term on the 
right-hand side
corresponds to the probability density that the particle is in $x$ at time $t$ having performed a free motion without resetting, occurring with probability $e^{-rt}$, 
while the second term refers to trajectories with last resetting at $t-\tau$ and no resetting events up to $t$, occurring with probability rate $re^{-r\tau}$, integrating over time and space \cite{Evans_2011,Evans_2018}. By using normalization condition the last integral is equal to $1$ and we recover equation (\ref{BM_ren}).

\medskip 
{Trapping.}
The results obtained above also apply to the case of trapping processes, consisting of  random
motion interrupted by sudden irreversible immobilization of particles at a given rate $r$.
By introducing the two PDFs $P_A$ of moving (active) particles and $P_B$ of immobilized (blocked) ones, we can write the kinetic equations
of the trapping process as
\begin{align}
\label{eqPA}
  \partial_t P_A(x,t) &= v^2 \partial_x^2 P_A(x,t) - r P_A(x,t) ,  \\
  \partial_t P_B(x,t) &= r  P_A(x,t) ,
\label{eqPB}
\end{align}
together with the initial condition $P_A(x,t=0)=\delta(x)$ and $P_B(x,t=0)=0$.
The last two terms in the preceding equations correspond to the sink and source terms for active and blocked particles, respectively, arising from the immobilization mechanism.
By differentiating with respect to time the equation for $P_B$, using that of $P_A$ and integrating over time we obtain
\begin{equation}
\label{eqPB2}
\partial_t P_B(x,t) = v^2 \partial_x^2 P_B(x,t) - r P_B(x,t) + r \delta(x)  , \\
\end{equation}
where we have considered the initial conditions.
We recognize that, by summing (\ref{eqPA}) and (\ref{eqPB2}), we obtain precisely the 
equation describing stochastic resetting (\ref{D_R}), valid for the total PDF  
$P(x,t)=P_A(x,t)+P_B(x,t)$, where the first term  represents mobile particles at time $t$, whereas the second term represents all immobilized particles at time $t$.
Alternatively, by applying Laplace and Fourier transforms to (\ref{eqPA})-(\ref{eqPB}),
is is easy to see that one obtains the expression (\ref{Pks_DR}) for the total PDF 
$\hat{\tilde P}$.
It is worth noting that, for the trapping process, equation (\ref{BM_ren}) has now the following intepretation:
the first term on the
right-hand side
is the probability density that the particle has not be trapped until time $t$ and 
the second one takes into account the particle's trapping at $x$ in a previous time $\tau$, occurring at rate $re^{-r\tau}$, 
and integrating over times.

\medskip
We then conclude that the function $\g$ given in (\ref{g_DR}) describes both diffusive motions 
with stochastic resetting and diffusive motions  in the presence of particles trapping.
As will be discussed later, this equivalence no longer holds for processes with memory.
We finally note that, since 
$\g(0)=r>0$, 
there exists a stationary PDF (\ref{Pst}), which has the following expression
\begin{equation}
    P_{st.}(x) = \frac{\sqrt{r}}{2v} \ \exp \left( -\frac{\sqrt{r}}{v} |x| \right) .
\end{equation}

\medskip
The MSD in the Laplace domain is
\begin{equation}
    \MSDLT = \frac{2v^2}{s(s+r)} = \frac{2v^2}{r} \left(\frac{1}{s}-\frac{1}{s+r} \right) ,
\end{equation}
which, in the time domain, reads
\begin{equation}
    \MSD = \frac{2v^2}{r} \left(1 - e^{-rt}\right) .
\end{equation}
For large times the MSD tends to a constant value 
\begin{equation}
    \MSD \to \frac{2v^2}{r} , \qquad t \to \infty ,
\end{equation}
related to the existence of a stationary distribution with a finite second moment.
At short times the particle performs a free diffusive motion
\begin{equation}
    \MSD \simeq 2v^2 t , \qquad t \to 0.
\end{equation}

\subsection{Anomalous Diffusion}
\label{Sec_AD}
Let us now consider a function $\g(s)$ which is a generic non-integer power of $s$
\begin{equation}
\label{g_AD}
\boxed{   
\g(s) = s^\mu 
}
\end{equation}
For 
$\sqrt{\g(s)}=s^{\mu/2}$ 
to be a Bernstein function, we must restrict the range of values 
of exponent to $0<\mu <2$ (see the Appendix).
In such a case we have 
\begin{equation}
\label{Pks_AD}
 {\hat {\tilde P}}(k,s) = \frac{s^{\mu-1}}{s^\mu+v^2 k^2} ,
\end{equation}
or, in a different form,
\begin{equation}
\label{ABM_1}
(s^\mu+v^2k^2) {\hat {\tilde P}}(k,s) = s^{\mu-1} .
\end{equation}
We  treat separately the two cases $0<\mu<1$ and $1<\mu<2$.\\

\noindent
For $0<\mu<1$ we can proceed as follows.
We rewrite the previous equation as
$$
s^{\mu-1} [s {\hat {\tilde P}}(k,s) - 1] = -v^2 k^2 {\hat {\tilde P}}(k,s) ,
$$
which, in the $(x,s)$ domain, becomes
$$
s^{\mu-1} [s {\tilde P}(x,s) - \delta(x)] = v^2 \partial_x^2 {\tilde P}(x,s) .
$$
By using the convolution theorem of Laplace transforms -- see (\ref{LT_conv}) --
we can write the previous equation in the $(x,t)$ domain as
\begin{equation}
    \int_0^t d\tau \ {\cal L}^{-1}[s^{\mu-1}](t-\tau) \ {\cal L}^{-1}[s {\tilde P}(x,s)-\delta(x)](\tau)
    = v^2 \partial_x^2 P(x,t) ,
\end{equation}
and, using (\ref{LT_der}) with $n=1$, (\ref{LT_pow}) and initial condition (\ref{ic}), we finally arrive at the following equation
\begin{equation}
 \frac{1}{\Gamma(1-\mu)} \int_0^t d\tau \ (t-\tau)^{-\mu} \ \partial_\tau P(x,\tau) = v^2 \partial_x^2 P(x,t) .
\end{equation}
We have then obtained a integro-differential equation with a power-law memory kernel. 
The term on the left is precisely what, in the literature, is called Caputo fractional derivative 
(see Appendix for basic definitions and references), 
and, therefore, we can write the equation of this random motion with memory in a concise form
\begin{equation}
\label{FDE}
    \partial_t^\mu P(x,t) = v^2 \partial_x^2 P(x,t) , \qquad \qquad  0<\mu<1 ,
\end{equation}
known as time-fractional diffusion equation \cite{Mainardi_1996}. 
It is worth noting how the introduction of the fractional Caputo derivative operator 
naturally appears as a consequence of the presence of a non-integer exponent $\mu$ in the expression of $\g(s)$ (\ref{g_AD}) appearing in the fundamental solution of the kinetic equation.
We observe that equation (\ref{FDE}) can be equivalently written in the following form
\begin{equation}
  \partial_t P(x,t) = v^2 D^{1-\mu} \partial_x^2 P(x,t) , \qquad \qquad  0<\mu<1 ,
\end{equation}
where $D^\mu$ is the Riemann-Liouville fractional derivative operator (see the Appendix).\\

\noindent
For $1<\mu<2$ we can rewrite (\ref{ABM_1}) as
$$
s^{\mu-2} [s^2 {\hat {\tilde P}}(k,s) - s] = -v^2 k^2 {\hat {\tilde P}}(k,s) ,
$$
and, proceeding as before, we finally arrive at the following equation
\begin{equation}
    \frac{1}{\Gamma(2-\mu)} \int_0^t d\tau \ (t-\tau)^{1-\mu} \ \partial_\tau^2 P(x,\tau)
    = v^2 \partial_x^2 P(x,t) ,
\end{equation}
which, again, can be written in the concise fractional form (see Appendix)
\begin{equation}
        \partial_t^\mu P(x,t) = v^2 \partial_x^2 P(x,t),  \qquad \qquad  1<\mu<2 ,
\end{equation}
which is known as fractional wave equation \cite{Mainardi_1996,Iafrate2020}.
Again we observe that the above equation can be written in a different form, 
in terms of the Riemann-Liouville  derivative operator
\begin{equation}
  \partial^2_t P(x,t) = v^2 D^{2-\mu} \partial_x^2 P(x,t) , \qquad \qquad  1<\mu<2 .
\end{equation}
\\

The solution of the fractional diffusion-wave equations, valid in the full range 
of exponent values $0<\mu<2$,  is given, in the Laplace-Fourier domain, by (\ref{Pks_AD}),
which reads in $(x,s)$, $(k,t)$ and $(x,t)$ domains
\begin{align}
 {\tilde P}(x,s) &= \frac{s^{\mu/2-1}}{2v} \exp\left( -\frac{|x|}{v}s^{\mu/2}\right) , \\
 {\hat P}(k,t) &= E_\mu (-v^2 k^2 t^\mu) , \label{Pkt_AD} \\
P(x,t) &= \frac{1}{2vt^{\mu/2}} M_{\mu/2}\left( \frac{|x|}{vt^{\mu/2}} \right) 
= \frac{1}{2v} \mathbb{M}_{\mu/2} \left( \frac{|x|}{v},t \right),
\label{Pxt_AD}
\end{align}
where $E_\mu(z)$ is the Mittag-Leffler function, $M_{\mu}(z)$ is the Mainardi (or $M$-Wright) function and $\mathbb{M}_\mu(x,t)$ is the $\mathbb{M}$-Wright function 
defined by $\mathbb{M}_\mu(x,t)=t^{-\mu} M_\mu(xt^{-\mu})$
(see the Appendix).
To derive (\ref{Pkt_AD}) from (\ref{Pks_AD}) we have used the property (\ref{LT_E1}).
The derivation of (\ref{Pxt_AD}) can be found in \cite{Mainardi_1996,Mainardi_2007,Mainardi_Rev_2010}, see also (\ref{Pxt_AD_M}) below.\\
The expression (\ref{Pxt_AD}) represents the general solution of the fractional diffusion-wave equation
valid for $0<\mu<2$.
By using known properties of the $\mathbb{M}$ function 
\cite{Mainardi_Rev_2010} we have that 
\begin{align}
\label{MW_1}
   & \mathbb{M}_{1/2} (|x|,t) = \frac{1}{\sqrt{\pi t}} \exp \left( - \frac{x^2}{4t} \right) , \qquad \qquad  \mu=1 ,  \\
    & \mathbb{M}_{1} (|x|,t) = \delta(x-t) + \delta(x+t)    ,\qquad \qquad \mu=2 .
   \label{MW_2}
\end{align}
thus obtaining for the PDF (\ref{Pxt_AD}), as limiting solutions,  the classical diffusive PDF (\ref{Pxt_D}) for $\mu=1$
and the wave-ballistic PDF (\ref{Pxt_B}) for $\mu=2$. \\

\noindent 
We now show how the stochastic processes described here, governed by time-fractional equations, 
can be associated to time-changed Brownian motions \cite{ISS_MS,meerschaert2019inverse}.
The $\mathbb{M}$-Wright function can be interpreted as a PDF of a time-changed process (see section \ref{Sec_Sub} below).
We consider a process that randomly 
associates to the {\it physical} time $t$ an {\it operational} time $u$ with a PDF $\mathbb{M}_\mu(u,t)$,
which depends on the parameter $\mu \in (0,1)$. 
In other words, the $\mathbb{M}$-Wright function can be viewed as the PDF of the inverse stable subordinator $E(t)=\inf \{u>0: D(u)>t\}$,
where $D(u)$ is a stable subordinator, i.e., a nondecreasing L\'evy process (stochastic process with stationary, independent increments),  
which satisfies the self-similarity property, $D(cu)$ has the same distribution of $c^{1/\mu} D(u)$  \cite{meerschaert2019inverse,applebaum2009levy}.
The solution of the fractional equations considered in
this section can be expressed as an average of a non-fractional diffusive process with respect to the time-changed PDF $\mathbb{M}_\mu$,
i.e., 
\begin{equation}
\label{PBMTC1}
    P(x,t) = \int_0^\infty du \ \mathbb{M}_\mu(u,t) \ P_{\mu=1}(x,u) ,  
\end{equation}
where $P_{\mu=1}$ is the solution of the classical diffusive equation, (\ref{Pxt_D}).
Indeed, by using the Laplace transform of the  $\mathbb{M}$-Wright function \cite{Mainardi_Rev_2010}
$$
{\mathcal L}[\mathbb{M}_\mu(u,t)] (u,s) = s^{\mu-1} e^{-us^\mu} , 
$$
we have that 
\begin{equation}
    {\hat{\tilde P}}(k,s) = s^{\mu-1} {\hat{\tilde P}}_{\mu=1}(k,s^\mu) = \frac{s^{\mu-1}}{s^\mu+v^2k^2} ,
\end{equation}
which is exactly the expression (\ref{Pks_AD}). 
Now, by noting that
$$
P_{\mu=1}(x,t) = \frac{1}{2v} \mathbb{M}_{1/2}\left( \frac{|x|}{v},t\right)
$$
we can write the PDF in the concise form (\ref{Pxt_AD}).
Indeed, we have
\begin{equation}
\label{Pxt_AD_M}
    P(x,t) = \frac{1}{2v} \int_0^\infty du \ \mathbb{M}_\mu(u,t) \ \mathbb{M}_{1/2}\left(\frac{|x|}{v},u\right)
 = \frac{1}{2v} \mathbb{M}_{\mu/2} \left( \frac{|x|}{v},t\right)
\end{equation}
where we have used the property -- see (4.23) in \cite{Mainardi_Rev_2010}
$$
\mathbb{M}_{\nu}(u,t) = \int_0^\infty d\tau \ \mathbb{M}_{\lambda}(u,\tau) \mathbb{M}_{\mu}(\tau,t) , \qquad \nu=\lambda \mu .
$$
We finally note that we could also have written the solution of the fractional equation as a time-changed 
process of the non-fractional wave solution 
\begin{equation}
\label{PBMTC2}
    P(x,t) = \int_0^\infty du \ \mathbb{M}_{\mu/2}(u,t) \ P_{\mu=2}(x,u) ,  
\end{equation}
with $P_{\mu=2}$ given by (\ref{Pxt_B}), where now the PDF of the time-changed process is given by the $\mathbb{M}$-Wright function of order $\mu/2$. 
It should be noted, however, that although the PDF $P(x,t)$ in (\ref{PBMTC1}) and (\ref{PBMTC2}) is the same, it corresponds to two distinct classes of underlying stochastic processes, obtained by applying different time-change mechanisms to the parent diffusive and wave solutions, respectively.
Take, for example, the limiting case $\mu=1$, corresponding to the heat kernel PDF (\ref{Pxt_D}).
Eq. (\ref{PBMTC2}) tells us that this PDF, besides corresponding to standard Brownian motion (Markovian process), is also obtained  by time-changing the wave solution (two propagating delta-peaks) with a direct process with inverse subordinator PDF given by $\mathbb{M}_{1/2}(u,t)$ (non-Markovian process). 
The differences between the underlying stochastic processes are not reflected in the one-time PDF, but rather in the corresponding multi-time joint probability densities \cite{Gardiner}.
\\

We now analyze the MSD and show why the above fractional equations describe anomalous behaviors.
In the full parameter range $0<\mu<2$ we have that the MSD is 
$$
\MSDLT = \frac{2v^2}{s^{1+\mu}} , 
$$
and, in the time domain, using (\ref{LT_pow}),
\begin{equation}
    \MSD = \frac{2v^2}{\Gamma(1+\mu)} t^\mu .
\end{equation}
Therefore, we obtain anomalous diffusive behaviors: subdiffusion (or slow-diffusion) for $0<\mu<1$ and 
superdiffusion (or fast-diffusion) for $1<\mu<2$.

\subsection{Anomalous diffusion with resetting}
\label{Sec_ADR}
We now consider the following form
\begin{equation}
\label{g_ADR}
\boxed{
    \g(s) = (s + r)^\mu 
} 
\end{equation}
In this case we have that $\sqrt{\g(s)}=(s+r)^{\mu/2} \in \mathcal{BF}$ when $0<\mu<2$,
being the composition of two Bernstein functions, $s^{\mu/2}$ and $s+r$ (see ({\it iii}) in the Appendix).
We have that
\begin{equation}
\label{Pks_ADR}
    {\hat {\tilde P}}(k,s) = \frac{1}{s} \frac{(s+r)^\mu}{(s+r)^\mu+v^2 k^2} = 
    \left( 1+\frac{r}{s} \right) {\hat {\tilde P}}_0(k,s+r) \ ,
\end{equation}
where the quantity $P_0$ refers to the PDF of the anomalous diffusion processes without resetting, see (\ref{Pks_AD}).
The inverse Fourier and Laplace transform of (\ref{Pks_ADR}) leads to
the renewal equation (\ref{BM_ren})
for stochastic resetting problems \cite{Evans_DSR,Evans_2018}.
Therefore, the process described by the function (\ref{g_ADR}), can be interpreted as an
anomalous diffusive random walk with stochastic resetting \cite{KGN_2019}.
It is worth noting that the kind of reset processes we are dealing with here 
are {\it hard} (or complete) resetting, involving not only particle position but
the entire process, thus erasing all memory effects.
The case of {\it soft} (or incomplete) resetting, in which only particle position is reset to the origin, 
requires a more accurate description, not captured by the simple form of the PDF given in (\ref{Pks}) 
\cite{KGN_2019,Shkilev_2022}.
\\
We now derive the equations governing this anomalous (hard) resetting process.
We start from (\ref{Pks_ADR}), rewritten as
$$
(s+r)^\mu {\hat {\tilde P}}(k,s) - \frac{(s+r)^\mu}{s} = -v^2k^2 {\hat {\tilde P}}(k,s) ,
$$
and apply the inverse Laplace operator, obtaining
\begin{equation}
\label{ILT1}
{\cal L}^{-1} \left[ s^\mu {\hat {\tilde P}}(k,s-r) - \frac{s^\mu}{s-r}\right] (k,t) 
= - v^2 e^{rt}  k^2 {\hat P}(k,t) ,
\end{equation}
where we have used the shift property (\ref{LT_trasl}).\\
Again we consider separately two cases, depending on the value of $\mu$.\\

\noindent 
We first consider $0<\mu<1$.
It is useful to write the term in the  left-hand side of (\ref{ILT1}) as
$$
\underbrace{
{\cal L}^{-1} \left[ 
s^{\mu-1} \left( 
s {\hat {\tilde P}}(k,s-r) -1 
\right)
\right]
}_{(1)}
-\underbrace{
{\cal L}^{-1} \left[ 
\frac{rs^{\mu-1}}{s-r} 
\right]}_{(2)} .
$$
The first term can be expressed, using the convolution theorem (\ref{LT_conv}) and the property (\ref{LT_pow}), as follows
$$
(1)= \frac{1}{\Gamma(1-\mu)}\int_0^t d\tau \ (t-\tau)^{-\mu} 
{\cal L}^{-1} \left[ 
s {\hat {\tilde P}}(k,s-r) -1 
\right] (\tau) = \partial_t^\mu \left(e^{rt} {\hat P}(k,t) \right) \ ,
$$
having considered that 
\begin{align*}
{\cal L}^{-1} \left[ s {\hat {\tilde P}}(k,s-r) -1 \right] (t) &=
{\cal L}^{-1} \left[ (s-r) {\hat {\tilde P}}(k,s-r)+r {\hat {\tilde P}}(k,s-r) -1 \right] (t) \\
&=e^{rt} \partial_t {\hat P}(k,t) + r e^{rt} {\hat P}(k,t) = \partial_t \left( e^{rt} {\hat P}(k,t)\right) ,
\end{align*}
and using the definition of fractional derivative 
(\ref{Capdev1}).
The second term, using as before the convolution theorem (\ref{LT_conv}) and the properties (\ref{LT_tn2}), (\ref{LT_pow}), is
$$
(2) = \frac{r}{\Gamma(1-\mu)}\int_0^t d\tau \ (t-\tau)^{-\mu} e^{r\tau} 
= r I^{1-\mu} e^{rt} ,
$$
where $I^\mu$ is the fractional integral 
(\ref{FracIntegral}).
Thus we have that the inverse Laplace-Fourier transform of (\ref{ILT1}) reads
\begin{equation}
\label{eq_ADR_1}
    \partial_t^\mu \left( e^{rt} P(x,t)\right) = v^2 e^{rt} \partial_x^2 P(x,t) +
    r \delta(x) I^{1-\mu} e^{rt} ,
\end{equation}
or, by noting that $\partial_t^\mu f(t) = I^{1-\mu} \partial_t f(t)$,
\begin{equation}
    I^{1-\mu} \left[ 
    e^{rt} \left( \partial_t P(x,t) + r P(x,t) -r \delta(x) \right) 
    \right] = v^2 e^{rt} \partial_x^2 P(x,t) .
\end{equation}
By applying the fractional derivative operator $D^{1-\mu}$ on both sides and using (\ref{FD_identity}), we can write the equation as
\begin{equation}
\label{FPE_ADR1}
    \partial_t P(x,t) = v^2 e^{-rt} D^{1-\mu} e^{rt} \partial_x^2 P(x,t) -rP(x,t) + r\delta(x) \ ,
\end{equation}
in agreement with the expression (13) reported in \cite{KGN_2019}.
We note that the operator appearing in the above equation, $e^{-rt} D^{1-\mu} e^{rt}$,
is a form of what is known in the literature as a truncated or tempered fractional derivative operator \cite{Meer_tempered,ALRAWASHDEH2017892,Garra202437}. 
It appears here as a consequence of the resetting process of the entire system, which involves both position and memory (renewal of the waiting time within the CTRW framework 
\cite{KGN_2019}.
\\

\noindent 
We now consider $1<\mu<2$.
In this case we can write the 
left-hand side
of (\ref{ILT1}) as
$$
\underbrace{
{\cal L}^{-1} \left[ 
s^{\mu-2} \left( 
s^2 {\hat {\tilde P}}(k,s-r) -s 
\right)
\right]
}_{(1)}
-\underbrace{
{\cal L}^{-1} \left[ 
\frac{rs^{\mu-1}}{s-r} 
\right]}_{(2)} .
$$
The first term gives
$$
(1)= \frac{1}{\Gamma(2-\mu)}\int_0^t d\tau \ (t-\tau)^{1-\mu} 
{\cal L}^{-1} \left[ 
s^2 {\hat {\tilde P}}(k,s-r) -s 
\right] (\tau) = \partial_t^\mu \left(e^{rt} {\hat P}(k,t) \right) +  \frac{r t^{1-\mu}}{\Gamma(2-\mu)}\ ,
$$
having 
applied the convolution theorem (\ref{LT_conv}) and (\ref{Capdev2}), and considered that 
\begin{align*}
{\cal L}^{-1} \left[ s^2 {\hat {\tilde P}}(k,s-r) -s \right] (t) &=
{\cal L}^{-1} \left[ (s-r)^2 {\hat {\tilde P}}(k,s-r)-(s-r) +r (2s-r) {\hat {\tilde P}}(k,s-r) -r \right] (t) \\
&=e^{rt} \partial_t^2 {\hat P}(k,t) + r e^{rt} {\cal L}^{-1} \left[ (2s+r){\hat {\tilde P}}(k,s) \right](t)
-r\delta(t) \\
&= e^{rt} \partial_t^2 {\hat P}(k,t) +
r e^{rt} {\cal L}^{-1} \left[ 2 (s{\hat {\tilde P}}(k,s)-1) +2+r{\hat {\tilde P}}(k,s)\right](t)
-r\delta(t) \\
&= e^{rt} \partial_t^2 {\hat P}(k,t) +
2 r e^{rt} \partial_t {\hat P}(k,t) +r^2 e^{rt} {\hat P}(k,t) +2r\delta(t) - r \delta(t) \\
&= \partial_t^2 \left( e^{rt} {\hat P}(k,t)\right) +r \delta(t) .
\end{align*}
For the second term we have
\begin{align*}
(2) &= r {\cal L}^{-1} \left[ s^{\mu-2}  \frac{s}{s-r}\right] (t)= 
\frac{r}{\Gamma(2-\mu)} \int_0^t d\tau (t-\tau)^{1-\mu} e^{r\tau} {\cal L}^{-1} \left[ \frac{s+r}{s}\right](\tau)\\
&=\frac{r}{\Gamma(2-\mu)} \int_0^t d\tau (t-\tau)^{1-\mu} e^{r\tau} \left[ \delta(\tau)+r\right]
= r^2 I^{2-\mu} e^{rt} + \frac{r t^{1-\mu}}{\Gamma(2-\mu)} ,
\end{align*}
where we have used the definition of the fractional integral (\ref{FracIntegral}).
Thus we have that the inverse Laplace-Fourier transform of (\ref{ILT1}) reads
\begin{equation}
\label{eq_ADR_2}
    \partial_t^\mu \left( e^{rt} P(x,t)\right) = v^2 e^{rt} \partial_x^2 P(x,t) +
    r^2 \delta(x) I^{2-\mu} e^{rt} ,
\end{equation}
or, by noting that $\partial_t^\mu f(t) = I^{2-\mu} \partial_t^2 f(t)$ (for $1<\mu<2$),
\begin{equation}
    I^{2-\mu} \left[ 
    e^{rt} \left( \partial_t^2 P(x,t) + 2r \partial_t P(x,t) + r^2 P(x,t) -r^2 \delta(x) \right) 
    \right] = v^2 e^{rt} \partial_x^2 P(x,t) .
\end{equation}
Finally, in terms of the operator $D^\mu$, the equation becomes
\begin{equation}
\label{FPE_ADR2}
    \partial_t^2 P(x,t) = v^2 e^{-rt} D^{2-\mu} e^{rt} \partial_x^2 P(x,t) 
    -2r \partial_t P(x,t) - r^2 P(x,t) + r^2 \delta(x) \ ,
\end{equation}
which extends the previous result to the case of fractional order derivative greater that one,
$1<\mu<2$.

\medskip
We note that we can write equations (\ref{eq_ADR_1}) and (\ref{eq_ADR_2}) in a unique form, 
valid in the full range of the exponent $0<\mu<2$.
Indeed, by defining the fractional integral $J^\mu$ as
\begin{equation}
\label{Jop}
J^\mu  = r^{\lceil \mu \rceil} I^{\lceil \mu \rceil-\mu} , 
\end{equation}
where $\lceil x \rceil$ is the ceiling function, giving the smallest integer greater than or equal to $x$,
we finally obtain the equation for anomalous diffusion with resetting in the concise form
\begin{equation}
\label{eq_ADR}
    \left( \partial_t^\mu - v^2 \partial_x^2 \right) e^{rt} P(x,t) =
    \delta(x) J^\mu e^{rt} . 
\end{equation}
We note that the last term can be also written as
\begin{equation}
    J^\mu e^{rt} = r I^{1-\mu} e^{rt} = r t^{1-\mu} E_{1,2-\mu}(rt) = 
    r^\mu  e^{rt} \frac{\gamma(1-\mu,rt)}{\Gamma(1-\mu)} ,
    \qquad {\text{for }} 0<\mu<1 ,
\end{equation}
and 
\begin{equation}
    J^\mu e^{rt} = r^2 I^{2-\mu} e^{rt} = r^2 t^{2-\mu} E_{1,3-\mu}(rt) = 
    r^\mu e^{rt} \frac{\gamma(2-\mu,rt)}{\Gamma(2-\mu)} ,
    \qquad {\text{for }} 1<\mu<2 ,
\end{equation}
where $\gamma(z,x)$ is the incomplete gamma function, $E_{\mu,\nu}(z)$ are the two-parameters Mittag-Leffler functions and we have used properties (\ref{LT_E2}) and (\ref{CFD_p2}) (see Appendix) and
the property (4.4.6) in \cite{gorenflo2020mittag}.
\\
\\
\noindent
Random motions with resetting are characterized by the existence of stationary PDFs.
In this case we have, from (\ref{Pst}),
\begin{equation}
    P_{st.}(x) = \frac{r^{\mu/2}}{2v} \ \exp \left( -\frac{|x|}{v} r^{\mu/2} \right) .
\end{equation}
The MSD is, in the Laplace domain,
\begin{equation}
\MSDLT = \frac{2v^2}{s (s+r)^\mu} .
\end{equation}
By using the property
$$
{\cal L}^{-1} \left[ \frac{1}{s (s+r)^\mu}\right] 
= e^{-rt} {\cal L}^{-1} \left[ \frac{1}{s^\mu (s-r)}\right] 
= e^{-rt} t^\mu E_{1,1+\mu}(rt) ,
$$
where $E_{\mu,\nu}$ is the two-parameters Mittag-Leffler function (see Appendix), 
we finally arrive a the expression of the MSD in the time domain \cite{KGN_2019}
\begin{equation}
\MSD = 2v^2 e^{-rt} t^\mu E_{1,1+\mu}(rt) = \frac{2v^2}{\Gamma(\mu)} \int_0^t d\tau \tau^{\mu-1} \ e^{-r\tau} = \frac{2v^2}{\Gamma(\mu)} r^{-\mu}\gamma(\mu,rt) ,
\label{MSD_ADr}
\end{equation}
where the second equality follows from the property (4.4.6) in \cite{gorenflo2020mittag} and $\gamma(z,x)$ is the incomplete gamma function (\ref{IGammaFun}).
Asymptotically the MSD is finite 
\begin{equation}
    \lim_{t \to \infty} \MSD = \lim_{s\to 0} s \MSDLT = \frac{2v^2}{r^\mu}, \qquad t \to \infty ,
\end{equation}
as also obtained from (\ref{MSD_ADr}) considering the limit $\lim_{t\to\infty} \gamma(\mu,t)=\Gamma(\mu)$.
The approach to the asymptotic value has the form $\MSD \simeq 2v^2r^{-\mu} [1 -(rt)^{\mu-1} e^{-rt}/\Gamma(\mu)]$, as obtained considering the asymptotic behavior of the incomplete gamma function for large values of $t$, $\gamma(\mu,t)\simeq \Gamma(\mu) -t^{\mu-1}e^{-t}$ \cite{gradshteyn2014table}.\\
At short time we have that
\begin{equation}
    \MSD \simeq \frac{2 v^2}{\Gamma(1+\mu)} t^{\mu} , \qquad t \to 0,
\end{equation}
as obtained from the Tauberian theorem (see Section \ref{Sec2}) or considering the expansion of the $\gamma$ function at small $t$,
$\gamma(\mu,t) \simeq  t^\mu/\mu - t^{1+\mu}/(1+\mu) + \cdots$ \cite{gradshteyn2014table}.


\subsection{Anomalous diffusion with trapping}
\label{Sec_ADT}
We now consider the function $\g$ of the form
\begin{equation}
\label{g_ADT}
\boxed{
    \g(s) = s^\mu + r 
}
\end{equation}
where $r>0$ and $0<\mu<1$ (in order $\sqrt{\g(s)}=\sqrt{s^\mu+r}$ to be a Bernstein function, using the properties 
of sum and composition of Bernstein functions, see ({\it i}) and ({\it iii}) in the Appendix).
The PDF (\ref{Pks}) now reads
\begin{equation}
\label{Pks_ADT}
    {\hat {\tilde P}}(k,s) = \frac{1}{s} \frac{s^\mu+r}{s^\mu+r+v^2 k^2}, 
\end{equation}
which leads to 
$$
(s^\mu+r+v^2k^2)  {\hat {\tilde P}}(k,s) = s^{\mu-1} +r s^{-1} ,
$$
or
$$
[s^\mu  {\hat {\tilde P}}(k,s) - s^{\mu-1} ] + r [ {\hat {\tilde P}}(k,s) - 1/s] = -v^2 k^2  {\hat {\tilde P}}(k,s) .
$$
By performing the inverse Fourier transform we obtain
\begin{equation}
s^\mu {\tilde P}(x,s) - s^{\mu-1} \delta(x) + r [{\tilde P}(x,s) - \delta(x)/s] = v^2 \partial_x^2 {\tilde P}(x,s) ,
\label{eq1AD_T}
\end{equation}
or also 
\begin{equation}
\label{eq2AD_T}
    s {\tilde P}(x,s)  -\delta(x) =\frac{s v^2 }{s^\mu+r} \partial_x^2 {\tilde P}(x,s) .
\end{equation}
By performing inverse Laplace transform of (\ref{eq1AD_T}), we finally arrive at the fractional-type equation
\begin{equation}
\label{eq_ADT}
\partial^\mu_t P(x,t)  = v^2 \partial_x^2 P(x,t) - r P(x,t)  + r \delta(x) ,
\end{equation}
which generalizes the diffusion equation with trapping (\ref{D_R}) to the case of anomalous diffusion.
Therefore, we conclude that the process described by the function $\g(s)$  (\ref{g_ADT}) is an anomalous diffusion process 
in the presence of trapping, i.e. with particle immobilization \cite{KOS_PRE2023} 
(cf. Eq. (\ref{eq2AD_T}) with Eq. (13) in  \cite{KOS_PRE2023}).\\
We note that the process can be also described starting from the fractional version of the equations
for the PDFs of moving (active, $P_A$) and immobilized (blocked, $P_B$) particles described in section \ref{Sec_DR}, 
leading to
\begin{align}
  \partial_t^\mu P_A(x,t) &= v^2 \partial_x^2 P_A(x,t) - r P_A(x,t) ,  \\
  \partial_t^\mu P_B(x,t) &= r  P_A(x,t) ,
\end{align}
which, in the Laplace-Fourier domain, gives precisely the expression (\ref{Pks_ADT})
for the total PDF $P=P_A+P_B$.\\
It is worth noting that, differently from the classical diffusion with trapping, the anomalous diffusion with 
trapping is described by a different function $\g$ with respect to the resetting case 
-- compare ({\ref{g_ADR}}) and (\ref{g_ADT}) -- resulting in different Fokker-Planck like equations,
(\ref{eq_ADR}) and (\ref{eq_ADT}).
Thus, memory effects play a role in determining different dynamics in the two cases.\\
The stationary state is described by the 
PDF (\ref{Pst}), which reads
\begin{equation}
    P_{st.}(x) = \frac{\sqrt{r}}{2v} \ \exp \left( -\frac{\sqrt{r}}{v} |x| \right) .
\end{equation}

Concerning the MSD we have
\begin{equation}
    \MSDLT = \frac{2v^2}{s(s^\mu+r)} = \frac{2v^2}{r} \left(\frac{1}{s}-\frac{s^{\mu-1}}{s^\mu+r} \right) ,
\label{MSDL_ADT}
\end{equation}
which, in the time domain, reads (see (\ref{LT_1sus}) and (\ref{LT_E1}) in the Appendix)
\begin{equation}
    \MSD = \frac{2v^2}{r} \left[ 1 - E_\mu(-rt^\mu)\right] ,
\label{MSD_ADT}
\end{equation}
where $E_\mu(z)$ is the Mittag-Leffler function. 
Asymptotically we have
\begin{equation}
    \lim_{t \to \infty} \MSD = \lim_{s\to 0} s \MSDLT = \frac{2v^2}{r}, \qquad t \to \infty ,
\end{equation}
approaching it as $\MSD\simeq (2v^2/r)[1-r^{-1}t^{-\mu}/\Gamma(1-\mu)]$, using the asymptotic expression of the Mittag-Leffler function,
$E_\mu(z) = -z^{-1}/\Gamma(1-\mu) + O(|z|^{-2})$ for $|z|\to \infty$ and $\Re(z)<0$
\cite{gorenflo2020mittag}. 
At short time, using Tauberian theorem, the MSD shows anomalous subdiffusive behavior 
\begin{equation}
    \MSD \simeq \frac{2 v^2}{\Gamma(1+\mu)} t^{\mu} , \qquad t \to 0 .
\end{equation}

\subsection{Run-and-tumble motion (telegraph equation)}
\label{Sec_RT}
We now consider the class of motions described by a function $\g(s)$  of the form
\begin{equation}
\boxed{
    \g(s)= s (s+\alpha)
}
\end{equation}
with $\alpha>0$.
In this case $\sqrt{\g(s)}=s^{1/2} (s+\alpha)^{1/2}$ is $\mathcal{CBF}$ (a subclass of $\mathcal{BF}$)
due to the fact that both $s$ and $s+\alpha$ are $\mathcal{CBF}$ and using 
properties ({\it ix}) reported in the Appendix.
We have 
\begin{equation}
\label{Pks_RT}
    {\hat {\tilde P}}(k,s) = \frac{s+\alpha}{s(s+\alpha)+v^2k^2} ,
\end{equation}
which can be recast in the form 
$$
s^2 {\hat{\tilde P}}(k,s) - s + \alpha \left[ s{\hat{\tilde P}}(k,s)-1\right] = -v^2 k^2 {\hat {\tilde P}}(k,s) .
$$
By performing inverse Fourier and Laplace transforms, using 
(\ref{FT_der}), (\ref{LT_der}) and considering initial conditions,
we finally obtain, in the $(x,t)$ domain
\begin{equation}
\label{eq_RT}
    \partial_t^2 P(x,t) + \alpha \partial_t P(x,t) = v^2 \partial_x^2 P(x,t) , 
\end{equation}
This is the so called {\it telegraph equation} (or telegrapher's equation),
which has a wide range of applications in many fields -- 
see, for example, Ref. \cite{weiss2002some} and references therein.
We just want to outline here that this equation describes one-dimensional run-and-tumble motions,
in which a particle moves along straight lines at constant speed $v$ and randomly reorients its direction
of motion at constant rate $\alpha$ (tumbling rate). 
Indeed, the kinetic equations for the 
PDF of right-moving ($P_R$) and left-moving ($P_L$) particles can be written as
\begin{align}
\label{RT_1d}
\partial_t P_R (x,t) &= -v \partial_x P_R (x,t)- \frac{\alpha}{2} P_R (x,t)+ \frac{\alpha}{2} P_L (x,t) ,\\  
\partial_t P_L (x,t)&= v \partial_x P_L (x,t)- \frac{\alpha}{2} P_L (x,t)+ \frac{\alpha}{2} P_R (x,t) , 
\end{align}
which, by considering initial conditions $P_R(x,t)|_{t=0}=P_L(x,t)|_{t=0}=\delta(x)/2$, are equivalent to the 
telegraph equation (\ref{eq_RT}) 
for the total PDF $P(x,t)=P_R(x,t)+P_L(x,t)$, 
with initial conditions $P(x,t=0)=\delta(x)$ and $\partial_t P(x,t)|_{t=0}=0$. 
Introduced to describe the motion of flagellated bacteria as {\it E.coli} 
\cite{Ecoli_Berg,berg1972,PhysRevE.48.2553,Berg_2022},
the run-and-tumble model has become one of the archetypal model for describing active particles motion
\cite{PhysRevLett.100.218103,Cates_2012,martens2012probability,angelani2013averaged,RevModPhys.88.045006}.
The fundamental solution of the telegraph equation is, in the Laplace domain, given by (\ref{Pxs})
\begin{equation}
    {\tilde P}(x,s) = \frac{1}{2v} \sqrt{\frac{s+\alpha}{s}} \exp \left( -\frac{|x|}{v} \sqrt{s(s+\alpha)}\right) .
\end{equation}
It the Fourier domain, the solution takes the form 
\begin{equation}
    {\hat P}(k,t) = e^{-\alpha t/2} \left( 
    \cosh{\frac{t}{2} \sqrt{\alpha^2-4 v^2 k^2}} +
    \frac{\alpha}{\sqrt{\alpha^2-4 v^2 k^2}} \sinh{\frac{t}{2} \sqrt{\alpha^2-4 v^2 k^2}}
    \right) ,
\end{equation}
as one can deduce by noting that, using  (\ref{LT_trasl}), the inverse Laplace transform of (\ref{Pks_RT}) can be written as
$$
{\mathcal L}^{-1}\left[ \frac{s+\alpha}{s(s+\alpha)+v^2k^2}\right] = 
e^{-\alpha t /2}  {\mathcal L}^{-1}\left[ \frac{s+\alpha/2}{s^2-\alpha^2/4+v^2 k^2}\right] ,
$$
and using the known results (\ref{LT_sh}) and  (\ref{LT_ch}).\\
Finally, in the $(x,t)$ domain, the PDF reads
\begin{equation}
P(x,t) =  \frac{e^{-\alpha t/2}}{2} \left[
\delta(x-vt)+\delta(x+vt)
+
\frac{\alpha}{2v}
\left(
I_0(\alpha \Delta /2v) +\frac{vt}{\Delta} I_1(\alpha \Delta/2v) 
\right) \ \theta(vt-|x|)
\right] .
\end{equation}
where $\Delta=\sqrt{v^2t^2-x^2}$, $\theta(x)$ is the Heaviside step function and 
$I_0$ and $I_1$ are modified Bessel functions  (\ref{MBF}) \cite{weiss2002some}.

\medskip
For the run-and-tumble motion the MSD reads
\begin{equation}
    \MSDLT = \frac{2v^2}{s^2(s+\alpha)} = 
    \frac{2v^2}{\alpha^2} \left( \frac{\alpha}{s^2} - \frac{1}{s}+  \frac{1}{s+\alpha}\right) ,
\end{equation}
which, in the time domain, gives the well known expression for active particles \cite{weiss2002some}
\begin{equation}
    \MSD = \frac{2v^2}{\alpha^2} \left[ \alpha t - 1 +e^{-\alpha t} \right].
\end{equation}
Diffusive and ballistic behaviors are observed at long and short times respectively
\begin{equation}
    \MSD \simeq \frac{2v^2}{\alpha} t , \  \qquad (t\gg \alpha^{-1}) , \qquad t \to \infty ,
\end{equation}
and
\begin{equation}
    \MSD \simeq v^2 t^2 , \  \qquad (t\ll \alpha^{-1}) , \qquad t \to 0.
\end{equation}

\subsection{Run-and-tumble motion with resetting or trapping}
\label{Sec_RTR}
We now consider
\begin{equation}
\label{g_RTR}
\boxed{
    \g(s) = (s+r) (s+r+\alpha) 
}
\end{equation}
The condition $\sqrt{\g(s)} \in \mathcal{CBF}$ is fulfilled following the same argument as in the previous case. 
However, now the third condition is verified in the form (\ref{c3b}), implying the existence of a stationary
asymptotic solution.
For this case the PDF reads
\begin{equation}
\label{Pks_RTR}
    {\hat {\tilde P}}(k,s) = \frac{1}{s} \ \frac{(s+r)(s+r+\alpha)}{(s+r)(s+r+\alpha)+v^2k^2} ,
\end{equation}
which can be rewritten as
$$
s^2 {\hat{\tilde P}}(k,s) - s + (\alpha+2r) \left[ s {\hat{\tilde P}}(k,s)-1\right] 
+r(\alpha+r) \left[ {\hat{\tilde P}}(k,s)-1/s \right] 
= -v^2 k^2 {\hat {\tilde P}}(k,s) .
$$
Using the known properties of Laplace and Fourier transforms (see Appendix) 
together with initial conditions (\ref{ic}) and (\ref{ic2}), 
the above equation, in the $(x,t)$ domain, corresponds to the second order differential equation
\begin{equation}
\label{eq_RTR}
    \partial_t^2 P(x,t) + (\alpha+2r) \partial_t P(x,t) + r (\alpha+r) P(x,t) 
    = v^2 \partial_x^2 P(x,t) 
    + r (\alpha+r)  \delta(x) ,
\end{equation}
which describes run-and-tumble motion in the presence of stochastic resetting.
Indeed, it is immediate to see that (\ref{Pks_RTR}) can be written in terms of the PDF
$P_0$ of the free RT model (\ref{Pks_RT}), 
leading to the renewal equations (\ref{BM_ren0}) and (\ref{BM_ren}), 
defining stochastic resetting \cite{Evans_DSR,Evans_2018}.

Similarly to the diffusive case, we can show that the function $\g$ (\ref{g_RTR}) and the corresponding 
equation (\ref{eq_RTR}) 
describe also RT motions in the presence of trapping \cite{Ang_PS2024}.
Indeed, assuming particle immobilization at constant rate $r$, we have that the PDF $P_A$ of moving particles
can be written in terms of PDF of free particles $P_0$ as
$$
P_A(x,t)= e^{-rt} P_0(x,t) ,
$$
and that of trapped ones $P_B$ obeys 
$$
\partial_t P_B(x,t) = r  P_A(x,t) .
$$
It is easy to see that the total PDF $P=P_A+P_B$ obeys the renewal equation (\ref{BM_ren}).\\
As in previous resetting or trapping cases there exits a stationary PDF, given by
\begin{equation}
    P_{st.}(x) = \frac{\sqrt{r(r+\alpha)}}{2v} \exp \left( -\frac{|x|}{v} \sqrt{r(r+\alpha)}\right) . 
\end{equation}
The MSD  in such a case is
\begin{equation}
    \MSDLT = \frac{2v^2}{s(s+r)(s+r+\alpha)} = 
    \frac{2v^2}{\alpha r (r+\alpha)} \left( \frac{r}{s+r+\alpha} +\frac{\alpha}{s} -  
    \frac{\alpha+r}{s+r}\right) ,
\end{equation}
which, in the time domain, gives
\begin{equation}
    \MSD = \frac{2v^2}{\alpha r (r+\alpha)} \left[ \alpha (1-e^{-rt})  - r e^{-rt} (1 -e^{-\alpha t}) \right].
\end{equation}
In the long time limit the MSD approaches a constant value
\begin{equation}
    \MSD \to  \frac{2v^2}{r(r+\alpha)},  \qquad t \to \infty , 
\end{equation}
while, at short times, we have ballistic behavior
$$
\MSD \simeq v^2 t^2 , \qquad t \to 0 .
$$
Possible diffusive behavior can be observed at intermediate times -- 
see, e.g., figure 3 in \cite{Ang_PS2024}, where  run-and-tumble motions have been investigated
in the presence of arbitrary extended trapping regions.

\subsection{Anomalous run-and-tumble motion}
\label{Sec_ART}
We now extend the case of run-and-tumble motion treated in \ref{Sec_RT} introducing non-integer powers
in the expression of the function $\g$
\begin{equation}
\boxed{
    \g(s) = s^\mu (s^\nu +\alpha) 
}
\end{equation}
Finding the range of values of exponents $\mu$ and $\nu$ for which 
$\sqrt{\g}$
is a Bernstein function needs subtle arguments.
The discussion is contained in a recent work \cite{Ang_Chaos2024}, 
but for the sake of completeness we also quote it here.
We can write for $0<\mu<2$
\begin{equation}
\label{eq_gs1}
\sqrt{\g(s)} = s^{\mu/2} (s^\nu+\alpha)^{1/2} = s^{\mu/2} h(s)^{1-\mu/2} ,
\end{equation}
where the function $h$ is
$$
h(s)=\left[ (s^\nu+\alpha) ^{1/\nu} \right]^{\nu/(2-\mu)} .
$$
Now, we observe that $s^{\nu/(2-\mu)} \in \mathcal{CBF}$ for $0<\nu<2-\mu$ and,
from property ({\it xii}) in the Appendix, we have that 
$(s^\nu+\alpha)^{1/\nu} \in  \mathcal{CBF}$ for $0<\nu<1$.
Therefore, using the composition of Bernstein functions ({\it viii}), we deduce that 
$h(s) \in \mathcal{CBF}$ for $0<\nu <\min(1,2-\mu)$.
From property ({\it ix}) applied to (\ref{eq_gs1}) we finally demonstrate that $\sqrt{\g(s)} \in \mathcal{CBF}$
for values of the exponents in the range (see also Figure 1 in \cite{Ang_Chaos2024} with 
$\epsilon=\nu$ and $\eta=\mu$)
\begin{equation}
\label{rangemunu}
    \{(\mu,\nu): 0<\mu<2, 0<\nu<\min(1,2-\mu)\} .
\end{equation}
Now  we analyze the PDF and deduce the Fokker-Planck equation of the underlying process.
We have that 
\begin{equation}
\label{Pks_ART}
    {\hat {\tilde P}}(k,s) = \frac{1}{s} \ \frac{s^\mu(s^\nu+\alpha)}{s^\mu(s^\nu+\alpha)+v^2k^2} ,
\end{equation}
and, then
$$
s^{\mu+\nu} {\hat{\tilde P}}(k,s) - s^{\mu+\nu-1} + \alpha \left[ s^\mu {\hat{\tilde P}}(k,s)-s^{\mu-1}\right] = -v^2 k^2 {\hat {\tilde P}}(k,s) .
$$
By using the property of Laplace transform of Caputo fractional derivative (\ref{CFD_p1}) 
and considering initial conditions (\ref{ic}) and (\ref{ic2}), 
we finally obtain the fractional telegraph equation for anomalous run-and-tumble motions
\begin{equation}
\label{eq_ART}
    \partial_t^{\mu+\nu} P(x,t) + \alpha \partial_t^{\mu} P(x,t) = v^2 \partial_x^2 P(x,t) .
\end{equation}
The reader can refer to Ref. \cite{Ang_JSP2024} (and references therein) for a recent exhaustive analysis 
of this equation and the corresponding random process in the case $\mu=\nu$, where it is shown that,
similarly to the case of anomalous Brownian motion (section \ref{Sec_AD}), also in this case we can describe the motion as a time-changed process (see also section \ref{Sec_Sub} below).
Investigations of  fractional kinetic equations based on different generalizations of 
the Cattaneo equation, 
which describes diffusion processes with finite velocity of propagation,
were reported in the seminal paper of Compte and Metzler \cite{CM1997},
extended to generalized kinetic equations with multiple parameters in 
Ref.s \cite{angelani2020fractional,Ang_Chaos2024}.\\
The explicit expression of Laplace transformed PDF  (\ref{Pxs}) reads
\begin{equation}
{\tilde P}(x,s) = \frac{\sqrt{s^{\mu-2}(s^\nu+\alpha)}}{2v} 
\exp \left( -\frac{|x|}{v}\sqrt{s^\mu(s^\nu+\alpha)}\right) .
\end{equation}
In the case of equal exponents $\mu=\nu$ (with $0<\mu<1$) we can also give an explicit 
expression of the PDF in the $(k,t)$ domain \cite{orsingher2004time,Ang_JSP2024,DGG2025}.
Indeed, in such a case we write (\ref{Pks_ART}) as
\begin{equation}
{\hat {\tilde P}}(k,s) = \frac{s^{\mu-1}(s^\mu+\alpha)}{s^{2\mu}+\alpha s^\mu +v^2k^2} 
= \frac{s^{\mu-1}(s^\mu+\alpha/2+\alpha/2)}{(s^\mu-s_1) (s^\mu - s_2)} ,
\end{equation}
where
$$
s_{1,2} = \frac{1}{2} \left( -\alpha \pm \sqrt{\alpha^2-4v^2 k^2}\right) .  
$$
Now we note that 
$$
\frac{s^\mu+\alpha/2}{(s^\mu-s_1) (s^\mu - s_2)} = \frac12 \left(\frac{1}{s^\mu-s_1} + \frac{1}{s^\mu-s_2} \right) ,
$$
and 
$$
\frac{\alpha/2}{(s^\mu-s_1) (s^\mu - s_2)} = \frac{\alpha/2}{\sqrt{\alpha^2-4v^2k^2}} \left(\frac{1}{s^\mu-s_1} - \frac{1}{s^\mu-s_2} \right) ,
$$
which, by using the Laplace transform (\ref{LT_E1}), leads to 
\begin{equation}
{\hat P}(k,t) = \frac12 \left[ 
\left( 1 +\frac{\alpha}{\sqrt{\alpha^2-4v^2k^2}}\right) E_\mu(s_1 t^\mu) +
\left( 1 -\frac{\alpha}{\sqrt{\alpha^2-4v^2k^2}}\right) E_\mu(s_2 t^\mu) 
\right] .
\end{equation}

The MSD is given by
\begin{equation}
\label{MSDL_ART}
    \MSDLT = \frac{2v^2}{s^{\mu+1}(s^\nu+\alpha)} ,
\end{equation}
which, in the time domain, is
\begin{equation}
    \MSD = 2v^2 t^{\mu+\nu} E_{\nu,1+\mu+\nu}(-\alpha t^\nu) , 
\end{equation}
where $E_{\mu,\nu}(z)$ is the two-parameters Mittag-Leffler function and we have used the property (\ref{LT_E2}) in the Appendix.
At long and short times the MSD is obtained using asymptotic behaviors of the Mittag-Leffler function, or, using the Tauberian theorem,
the small and large $s$ expansion of (\ref{MSDL_ART}), obtaining 
\begin{equation}
    \MSD \simeq \frac{2v^2}{\alpha \Gamma(1+\mu)} t^\mu , \qquad  t \to \infty ,
\end{equation}
and 
\begin{equation}
    \MSD \simeq \frac{2v^2}{\Gamma(1+\mu+\nu)} t^{\mu+\nu} , \qquad t \to 0 .
\end{equation}
Note that the asymptotic (long-time) behavior is determined only by the exponent $\mu$,
resulting in a subdiffusive motion for $\mu<1$ and superdiffusive for $1<\mu<2$.

\subsection{Anomalous run-and-tumble motion with resetting}
\label{Sec_ARTR}
We now generalize the previous model by considering
\begin{equation}
\boxed{
    \g(s) = (s+r)^\mu ((s+r)^\nu+\alpha) 
}
\end{equation}
We can repeat the arguments of the previous section and deduce that the range of values of the exponents 
ensuring that $\sqrt{\g}$ is a Bernstein function is 
$\{(\mu,\nu): 0<\mu<2, 0<\nu<\min(1,2-\mu)\}$.
The expression for the PDF reads
\begin{equation}
\label{Pks_ARTR}
    {\hat {\tilde P}}(k,s) = \frac{1}{s} \frac{(s+r)^\mu((s+r)^\nu+\alpha)}{(s+r)^\mu((s+r)^\nu+\alpha)+v^2 k^2} .
\end{equation}
It is immediate to see that this can be rewritten in terms of the PDF $P_0$ of the
anomalous run-and-tumble motion (\ref{Pks_ART}), 
resulting in the renewal equations  
(\ref{BM_ren0}) and (\ref{BM_ren})
for stochastic hard resetting \cite{Evans_DSR,Evans_2018,KGN_2019,Gor_2024,Sandev_2024}, 
in analogy with the case of anomalous diffusion with resetting, see section \ref{Sec_ADR} and discussion after (\ref{Pks_ADR}).
Proceeding as in the previous sections, we can write 
$$
(s+r)^{\mu+\nu} \left[ {\hat {\tilde P}}(k,s)-\frac{1}{s}\right] +
\alpha (s+r)^\mu \left[ {\hat {\tilde P}}(k,s)-\frac{1}{s}\right] 
= -v^2k^2 {\hat {\tilde P}}(k,s) ,
$$
and apply the inverse Laplace operator, 
\begin{equation}
{\cal L}^{-1} \left[ Y_{\mu+\nu}(k,s) \right] (k,t) +
\alpha {\cal L}^{-1} \left[ Y_{\mu}(k,s) \right] (k,t) 
= - v^2 k^2 {\hat P}(k,t) ,
\end{equation}
where 
\begin{equation}
    Y_{\mu}(k,s) = (s+r)^\mu \left[ {\hat {\tilde P}}(k,s)-\frac{1}{s}\right]  .
\end{equation}
The inverse Laplace transform of $Y_\mu$ is, for $0<\mu<1$
\begin{equation}
e^{rt} {\cal L}^{-1} \left[ Y_{\mu}(k,s) \right] (k,t) =
 \partial_t^\mu \left( e^{rt} {\hat P}(k,t) \right) - r I^{1-\mu} e^{rt} , 
\end{equation}
and, for $1<\mu<2$
\begin{equation}
e^{rt} {\cal L}^{-1} \left[ Y_{\mu}(k,s) \right] (k,t) =
\partial_t^\mu \left( e^{rt} {\hat P}(k,t) \right) - r^2 I^{2-\mu} e^{rt} ,
\end{equation}
as obtained in Section \ref{Sec_ADR}.
Therefore, we can distinguish three zones.\\
In the region 
$\{(\mu,\nu): 0<\mu<1, 0<\mu+\nu<1\}$ we have 
\begin{equation}
  \left( \partial_t^{\mu+\nu} + \alpha \partial_t^\mu \right)  e^{rt} P(x,t)  
  = v^2 e^{rt} \partial_x^2 P(x,t) + r \delta(x) \left( I^{1-\mu-\nu}+\alpha I^{1-\mu}\right) e^{rt} .
\end{equation}
In the region 
$\{(\mu,\nu): 0<\mu<1, 1-\mu<\nu<1\}$ (where $1<\mu+\nu<2$) we obtain the equation
\begin{equation}
  \left( \partial_t^{\mu+\nu} + \alpha \partial_t^\mu \right)  e^{rt} P(x,t)  
  = v^2 e^{rt} \partial_x^2 P(x,t) + r \delta(x) \left( r I^{2-\mu-\nu}+\alpha I^{1-\mu}\right) e^{rt} .
\end{equation}
Finally, considering 
$\{(\mu,\nu): 1<\mu<2, 0<\nu<2-\mu\}$, where both $\mu$ and $\mu+\nu$ are in the interval $(1,2)$, we obtain
\begin{equation}
  \left( \partial_t^{\mu+\nu} + \alpha \partial_t^\mu \right)  e^{rt} P(x,t)  
  = v^2 e^{rt} \partial_x^2 P(x,t) + r^2 \delta(x) \left(  I^{2-\mu-\nu}+\alpha I^{2-\mu}\right) e^{rt} .
\end{equation}
The above equations can be written in a more concise form using the fractional integral operator $J^\mu$ defined in (\ref{Jop}),
obtaining the general equation for anomalous run-and-tumble motion with resetting, 
valid in the whole parameters range
\begin{equation}
    \left( \partial_t^{\mu+\nu} + \alpha \partial_t^\mu - v^2 \partial_x^2 \right) e^{rt} P(x,t) 
    = \delta(x) \left( J^{\mu+\nu}+ \alpha J^\mu \right) e^{rt} . 
\end{equation}
By applying (\ref{CIrel}) we can write the above equations in similar forms obtained for the anomalous diffusion cases, (\ref{FPE_ADR1}) and (\ref{FPE_ADR2}).\\

The PDF in the $(x,s)$ domain is
\begin{equation}
    {\tilde P}(x,s) = \frac{\sqrt{(s+r)^\mu((s+r)^\nu+\alpha)}}{2vs} 
    \exp \left( - \frac{|x|}{v} \sqrt{(s+r)^\mu((s+r)^\nu+\alpha)}\right) ,
\end{equation}
and the expression of the stationary solution reads in this case
\begin{equation}
    P_{st.}(x) = \frac{\sqrt{r^\mu(r^\nu+\alpha)}}{2v} \exp \left( -\frac{|x|}{v} \sqrt{r^\mu(r^\nu+\alpha)}\right) . 
\end{equation}
The MSD is, in the Laplace domain,
\begin{equation}
\MSDLT = \frac{2v^2}{s (s+r)^\mu ((s+r)^\nu+\alpha)} .
\end{equation}
By performing the inverse Laplace transform, using (\ref{LT_int}) and (\ref{LT_trasl}), we obtain
$$
\mathcal{L}^{-1}[\MSDLT](t) = 2v^2 \int_0^t d\tau\ e^{-r\tau} \mathcal{L}^{-1}[\frac{1}{s^\mu(s^\nu+\alpha)}](\tau),
$$
which, using (\ref{LT_E2}), finally leads to
\begin{equation}
    \MSD = 2v^2 \int _0^t d\tau \ e^{-r\tau} \tau^{\mu+\nu-1} E_{\nu,\mu+\nu}(-\alpha \tau^\nu) .
\end{equation}
In the long time limit the MSD tends to a constant
\begin{equation}
    \MSD \to \frac{2v^2}{r^\mu(r^\nu+\alpha)} , \qquad t \to \infty ,
\end{equation}
while, at short times, using the Tauberian theorem, we have
\begin{equation}
    \MSD \simeq \frac{2v^2}{\Gamma(1+\mu+\nu)} \ t^{\mu+\nu}, \qquad t \to 0 .
\end{equation}

\subsection{Anomalous run-and-tumble motion with trapping}
\label{Sec_ARTT}
We conclude by discussing the following form of $\g(s)$
\begin{equation}
\boxed{
    \g(s) = (s^\mu+r) (s^\nu+r+\alpha)
}
\end{equation}
The range of values of exponents for which  $\sqrt{\g} \in \mathcal{BF}$ is $0<\mu<1$, $0<\nu<1$ 
-- see the discussion in section \ref{Sec_ADT} and property ({\it ix}) in the Appendix.
The PDF reads
\begin{equation}
    {\hat {\tilde P}}(k,s) = \frac{1}{s} \ \frac{(s^\mu+r)(s^\nu+r+\alpha)}{(s^\mu+r)(s^\nu+r+\alpha)+v^2k^2} ,
\end{equation}
which can be rewritten as
$$
\left[ s^{\mu+\nu} + (r+\alpha) s^\mu +r s^\nu + r (r+\alpha)  \right]
\left( {\hat {\tilde P}}(k,s)-\frac{1}{s} \right)
= -v^2 k^2 {\hat {\tilde P}}(k,s) .
$$
The above equation, in the $(x,t)$ domain, corresponds to
-- using properties (\ref{CFD_p1}), (\ref{LT_1sus}), (\ref{FT_der}) and (\ref{FT1}) in the Appendix  
\begin{equation}
\label{eq_ARTT}
\left[    \partial_t^{\mu+\nu} + (r+\alpha) \partial_t^\mu 
+ r \partial_t^\nu+ r (\alpha+r) \right]P(x,t) 
    = v^2 \partial_x^2 P(x,t) 
    + r (\alpha+r)  \delta(x) ,
\end{equation}
which generalizes the trapping equation (\ref{eq_RTR}) to the case of anomalous motion.
The above equation (\ref{eq_ARTT}) thus describes anomalous run-and-tumble walks
in the presence of particles trapping.
For $\mu=\nu=1$ it reduces to (\ref{eq_RTR}), while for $r=0$ we obtain the anomalous run-and-tumble 
equation (\ref{eq_ART}).
We also note that, in the diffusive limit $v,\alpha \to \infty$ with $v^2/\alpha \sim const.$,
we recover the anomalous diffusion equation with trapping, (\ref{eq_ADT}).\\
The Laplace transformed PDF has the form
\begin{equation}
    {\tilde P}(x,s) = \frac{\sqrt{(s^\mu+r)(s^\nu+r+\alpha)}}{2vs} 
    \exp \left( - \frac{|x|}{v} \sqrt{(s^\mu+r)(s^\nu+r+\alpha)}\right) ,
\end{equation}
and the stationary PDF reads
\begin{equation}
    P_{st.}(x) = \frac{\sqrt{r(r+\alpha)}}{2v}
    \exp \left( - \frac{|x|}{v} \sqrt{r(r+\alpha)}\right) .
\end{equation}

For the MSD we have
\begin{equation}
    \MSDLT = \frac{2v^2}{s(s^\mu+r)(s^\nu+r+\alpha)} ,
\end{equation}
which can be written, in the time domain, as integral of Mittag-Leffler functions
\begin{equation}
    \MSD = \frac{2v^2}{r+\alpha} \int_0^t d\tau \ 
    (t-\tau)^{\mu-1} E_{\mu,\mu}\left(-r(t-\tau)^\mu \right)
    \left[ 1 - E_\nu\left(-(r+\alpha)\tau^\nu\right)\right] ,
\end{equation}
where we have used (\ref{LT_conv}) with $f=(s^\mu+r)^{-1}$ and $g=(s(s^\nu+r+\alpha))^{-1}$, (\ref{LT_E2}) and the results in (\ref{MSDL_ADT}) and (\ref{MSD_ADT}).
Asymptotically we have that, independently of the values of the exponents $\mu$ and $\nu$,
the MSD approaches a constant value
\begin{equation}
    \MSD \to \frac{2v^2}{r(r+\alpha)} ,\qquad t \to \infty .
\end{equation}
At short times instead we have 
\begin{equation}
    \MSD \simeq \frac{2v^2}{\Gamma(1+\mu+\nu)} \ t^{\mu+\nu}, \qquad t \to 0.
\end{equation}

\section{Subordinators, parent processes and time changing}
\label{Sec_Sub}
In the analysis of anomalous diffusion in Section \ref{Sec_AD}, we have shown that the process can be viewed as
a time-changed Brownian motion. In this Section, we try to generalize this approach, aiming to understand how the various processes analyzed in this work can be described in terms of some time-changed parent process. 
In such a case the PDF, solution of the generalized Fokker-Planck equation  (\ref{FPeq}), can be written as an integral transformation \cite{ISS_MS,meerschaert2019inverse}
\begin{equation}
\label{P_child}
    P(x,t) = \int_0^\infty du\ h(u,t) P_0(x,u) ,
\end{equation}
where $P_0$ is the PDF of the parent process and $h(u,t)$ is the probability density of the time-changing process (directing process), which associates an operational time $u$ to the physical time $t$. The function $h$ can be viewed as the PDF of an inverse subordinator.\\ 
\noindent
We briefly recall some key terminology \cite{applebaum2009levy,schilling-bern,Bertoin1999}. \\
\noindent
A {\it L\`evy process} is a stochastically continuous process starting from zero and with independent, stationary increments.\\
\noindent
A {\it subordinator} $D=\{D(u), u\geq 0\}$ is a non-negative and non-decreasing L\`evy process.
Let $g(t,u)$ be the PDF of the subordinator $D$. 
The Laplace transform $\tilde{g}(s,u)$ of the distribution $g(t,u)$ is given by
\begin{equation}
    \mathbb{E}[e^{-sD(u)}] = \tilde{g}(s,u)=e^{-u \psiD(s)} ,
\label{LT_sub}
\end{equation}
where $\psiD(s)$ is called the {\it Laplace exponent} of the subordinator $D$ 
and $\mathbb{E}$ represents the expectation value, i.e., $\mathbb{E}[f(X)] = \int f(x) p_X(x) dx$ where $X$ is a random variable with PDF $p_X(x)$.
The Laplace exponent is a Bernstein function with $\psiD(0)=0$ and 
has the representation
\cite{applebaum2009levy,schilling-bern}
\begin{equation}
    \psiD(s) = b s+\int_0^\infty(1-e^{-st}) \mu(dt) , 
\label{LKf}
\end{equation}
where $\mu(dt)=m(t) dt$ is the L\`evy measure of $D$, 
$b$ the drift coefficient and the integral converges, i.e., $\int_0^\infty (1 \wedge t) \mu(dt)<\infty$, where $a\wedge b=\min\{a,b\}$. \\
\noindent 
The {\it inverse subordinator} $E=\inf\{u>0: D(u)>t\}$ is the first passage time of the subordinator $D$ above $t\geq0$.
We observe that $\mathbb{P}[E(t)\leq u] = \mathbb{P}[D(u) \geq t] = \int_t^\infty g(\tau,u) d\tau$ and then 
the PDF of the inverse subordinator $h(u,t)=\frac{d}{du} \mathbb{P}[E(t)\leq u]$ has the Laplace transform \cite{meerschaert2008triangular}
\begin{equation}
    \tilde{h}(u,s) = \frac{\psiD(s)}{s} e^{-u\psiD(s)} .
\label{hus}
\end{equation}
Given a subordinator $D$ with Laplace exponent $\psiD$, we define the {\it killed subordinator} $D_k=\{D_k(u),u\geq 0\}$ by the prescription \cite{applebaum2009levy,schilling-bern}
\begin{equation}
  D_k(u)=\begin{cases}
    D(u), & \text{for $0\leq u<T$} ,\\
    \infty, & \text{for $u\geq T$}, 
  \end{cases}
\end{equation}
where $T$ is an exponentially distributed random variable (independent on $D$) with PDF $p_T(x)=a e^{-ax}$. 
The process $D_k$ is the subordinator $D$ killed at independent exponential times.
There is a one-to-one correspondence between killed subordinators $D_k$ and Bernstein functions $\psiDk$ \cite{applebaum2009levy},
given by
\begin{equation}
    \mathbb{E}[e^{-sD_k(u)}] =e^{-u \psiDk(s)} ,
\label{LT_subk}
\end{equation}
where the function $\psiDk(s)$, generalization of (\ref{LKf}), is $\psiDk(s)=a+\psiD(s)$ and, expressed in the L\`evy-Khintchine representation, is
    \begin{equation}
    \psiDk(s) = a+ b s+\int_0^\infty(1-e^{-st}) m(t) dt , 
\label{LKf2}
\end{equation}
where $a$ is called the killing rate (for $a>0$ the stochastic process has a finite lifetime \cite{schilling-bern,Bertoin1999}). \\
The {\it inverse of the killed subordinators} $D_k$ is $E_k=\inf\{u>0: D_k(u)>t\}$. 
By observing that $\mathbb{P}[E_k(t)\leq u] = 1-e^{-au} + e^{-au} \mathbb{P}[E(t) \leq u]$ we have that the PDF of the inverse subordinator $E_k$ can be written as
\begin{equation}
    \tilde{h}(u,s) = \frac{\psiDk(s)}{s} e^{-u\psiDk(s)} .
\label{husk}
\end{equation}

\noindent 
In conclusion, considering a general (possibly killed) subordinator, we can write the PDF $P$ (\ref{P_child}) of the considered process, in the Fourier-Laplace domain, in terms of the parent process $P_0$ and the Laplace exponent $\psiDk$ of the subordinator as follows
\begin{equation}
\label{PkspsiD}
    \hat{\tilde{P}}(k,s) = \frac{\psiDk(s)}{s} \hat{\tilde{P}}_0(k,\psiDk(s)) .
\end{equation}

\medskip
{\bf Parent Brownian process.}
By taking the Brownian motion as the parent process, whose PDF in $(k,s)$ domain is given by (\ref{PDF_ks_BM})
$$
\hat{\tilde P}_0(k,s) = \frac{1}{s+v^2 k^2} , 
$$
we have that (\ref{PkspsiD}) reads
\begin{equation}
    \hat{\tilde{P}}(k,s) = \frac1s \frac{\psiDk(s)}{\psiDk(s)+v^2k^2} .
\end{equation}
By comparing with (\ref{Pks}) we have $\g(s)=\psiDk(s)$. However, $\psiDk(s)$ must now be a Bernstein function, rather than merely its square root (\ref{c1}). 
This restriction limits the class of processes that can be obtained by subordinating the parent Brownian process.
In particular we have the following correspondence between $\psiDk$ and the  different subordinated processes
\begin{itemize}
\item[--] $\psiDk(s)=s+r$ corresponds to diffusion with resetting or trapping;
\item[--] $\psiDk(s)=s^\mu$ with $0<\mu<1$ to anomalous diffusion;
\item[--] $\psiDk(s)=(s+r)^\mu$ with $0<\mu<1$ to anomalous diffusion under resetting;
\item[--] $\psiDk(s)=s^\mu+r$ with $0<\mu<1$ to anomalous diffusion with trapping.
\end{itemize}
We note that only processes with parameter $\mu<1$ (subdiffusive models) are allowed, i.e., can be considered as subordinated processes of a parent Brownian process.

\medskip
{\bf Parent run-and-tumble process.}
 Taking the run-and-tumble model as the parent process (\ref{Pks_RT})
$$
\hat{\tilde P}_0(k,s) = \frac{s+\alpha}{s(s+\alpha)+v^2 k^2} , 
$$
we have 
\begin{equation}
    \hat{\tilde{P}}(k,s) = \frac1s \frac{\psiDk(s) (\psiDk(s)+\alpha)}{\psiDk(s)(\psiDk(s)+\alpha)+v^2k^2} .
\end{equation}
By comparing with (\ref{Pks}) we have that $\g(s)=\psiDk(s) (\psiDk(s)+\alpha)$. 
Similarly to the previous case, we have that the processes that can be obtained by subordinating the parent run-and-tumble process have the following correspondence with the Laplace exponent:
\begin{itemize}
    \item[--] $\psiDk(s)=s+r$ corresponds to the RT motion with resetting or trapping;
\item[--] $\psiDk(s)=s^\mu$ with $0<\mu<1$ to the anomalous RT motion;
\item[--] $\psiDk(s)=(s+r)^\mu$ with $0<\mu<1$ to the anomalous RT motion under resetting;
\item[--] $\psiDk(s)=s^\mu+r$ with $0<\mu<1$ to the anomalous RT motion with trapping. 
\end{itemize} 
As before, we note that only processes with parameter $\mu<1$ are allowed, and,
with respect to the cases analyzed in previous sections where two parameters are present, we have to consider only $\nu=\mu<1$.

\medskip
It is worth noting that, considering the ballistic motion as parent process, we obtain that $\psiDk(s)=\sqrt{\g(s)}$, and, therefore, all the cases treated in this manuscript (for which $\sqrt{\g(s)}$ is a Bernstein function) can be considered subordinated processes with respect to the ballistic motion, which represents a superposition of constant speed motions along the positive and negative direction of the axis. In such a case the characteristic of the different kinds of motion are encoded entirely in the directing process of time-changing, while the parent process only takes into account the upward and downward direction of the motion along the $x$-axis (see for example the discussion in \cite{Ang_Chaos2024}).

\medskip
In conclusion, the different types of random motion analyzed in this work can be obtained as subordinated processes of standard Brownian or run-and-tumble motions (at least for the cases for which $0<\mu<1$), with only four kinds of subordinators. 

\begin{enumerate}
    \item  $ \boxed{\psi_{1}(s)=s+r}$ (killed pure drift subordinator). \\
    The PDF of the inverse subordinator is, in the Laplace domain,  $\tilde{h}_1(u,s)=(1+r/s)e^{-u(r+s)}$, and in the time domain
$h_1(u,t)=e^{-ur} \delta(t-u)+re^{-ur} \theta(t-u)$. It it easy to show that, substituting in (\ref{P_child}), we obtain exactly the renewal equation for the resetting process (\ref{BM_ren}) \cite{GBM_Stoj2021}.

        \item $\boxed{\psi_{2}(s)=s^\mu}$ ($\mu$-stable subordinator, $0<\mu<1$).\\
In this case the Laplace exponent can be written as
\begin{equation}
    \psi_{2}(s)= \int_0^\infty (1-e^{-st}) \frac{\mu}{\Gamma(1-\mu)} t^{-1-\mu} dt .
\end{equation}
Indeed, by writing $1-e^{-st}=t\int_0^s e^{-xt} dx$, we have $\psi_2(s)=(\mu/\Gamma(1-\mu)) \int_0^s {\mathcal L}[t^{-\mu}](x) dx$ and, using (\ref{LT_pow}) we finally obtain the previous expression.  
The PDF of the inverse stable subordinator has the form $\tilde{h}_2(u,s)=s^{\mu-1} e^{-us^\mu}$.

\item $\boxed{\psi_{3}(s)=(s+r)^\mu}$ (killed tempered $\mu$-stable subordinator, $0<\mu<1$).\\
Now we have that 
\begin{equation}
    \psi_{3}(s)= r^{\mu}+\int_0^\infty (1-e^{-st}) \frac{\mu}{\Gamma(1-\mu)} t^{-1-\mu} e^{-rt}dt ,
\end{equation}
as obtained by following steps similar to the previous ones, leading to 
$\psi_3(s)-r^\mu=(\mu/\Gamma(1-\mu)) \int_0^s {\mathcal L}[t^{-\mu}](x+r) dx$ and using (\ref{LT_pow}) and (\ref{LT_trasl}).
The PDF of the inverse subordinator reads $\tilde{h}_3(u,s)=s^{-1}(s+r)^{\mu} e^{-u(s+r)^\mu}$. 
We observe that $\psi_3$ is the killed version (with killing rate $r^\mu$) of the Laplace exponent of the so called tempered $\mu$-stable subordinator,  $(s+r)^\mu-r^\mu$ \cite{ALRAWASHDEH2017892,Kumar2015,Meer_tempered}.\\
It is interesting to note that the Laplace exponent can be written as a composite function of the previous two exponents, 
$\psi_3(s)=\psi_2(\psi_1(s))=(\psi_2 \circ \psi_1)(s)$, and the corresponding PDF of the inverse subordinator has the form
$$
h_3(u,t) = \int_0^\infty h_1(v,t) h_2(u,v) dv ,
$$
as can be easily verified by considering the Laplace transform and the expression (\ref{husk}).
The PDF $h_3$ can then be viewed as a subordinated process obtained by time-changing the PDF $h_2$ with the directing process $h_1$. 
By using the explicit expression of $h_1$ we obtain
$$
h_3(u,t)= e^{-rt} h_2(u,t) + r \int_0^t e^{-rv} h_2(u,v) dv ,
$$
which is the renewal equation for the PDF of the inverse $\mu$-stable subordinator.
Therefore, the subordinated process $P$ can be obtained applying two time-changing to the parent process $P_0$, first $h_2$ and then $h_1$
$$
P(x,t) =  \int_0^\infty h_3(u,t) P_0(x,u) du = \int_0^\infty h_1(v,t) \left( \int_0^\infty h_2(u,v) P_0(x,u) du \right) dv .
$$

\item $\boxed{\psi_{4}(s)=s^\mu + r}$ (killed $\mu$-stable subordinator, $0<\mu<1$).\\
Now we have 
\begin{equation}
    \psi_{4}(s)= r+\int_0^\infty (1-e^{-st}) \frac{\mu}{\Gamma(1-\mu)} t^{-1-\mu} dt 
\end{equation}
and $\tilde{h}_4(u,s)=s^{-1}(s^\mu+r) e^{-u(s^\mu+r)}$.
In this case we have $\psi_4(s)=\psi_1(\psi_2(s))=(\psi_1 \circ \psi_2)(s)$, in the opposite order with respect to the previous case.
Now we get
$$
h_4(u,t) = \int_0^\infty h_2(v,t) h_1(u,v) dv ,
$$
or, using the expression of $h_1$
$$
h_4(u,t)= e^{-ru} h_2(u,t) + r e^{-ru} \int_u^\infty  h_2(v,t) dv .
$$
We have that the generic subordinated process $P$ is obtained first time-changing the parent process $P_0$ with $h_1$ and then with $h_2$
$$
P(x,t) = \int_0^\infty h_2(v,t) \left( \int_0^\infty h_1(u,v) P_0(x,u) du \right) dv .
$$

\end{enumerate}

\section{Conclusions}
\label{Sec_Concl}
In this paper we have analyzed in a unified way random motions in one dimensional space
described by generic integro-differential equations whose fundamental solutions 
take a simple form in terms of a function $\g(s)$, see (\ref{Pks}).
The one-dimensional case examined here admits an exact analytical treatment while providing, in many situations, a reliable approximation of more complex processes whose dynamics can be effectively captured by a single spatial degree of freedom.
All the properties of the different motions are encoded in the function $\g(s)$,
which must meet some conditions in order to describe a random
motion process, summarized in Table I. 
Reviewing all possible simple functions $\g(s)$ that satisfy the required conditions
(exploiting fundamental properties of completely monotonic and Bernstein functions),
we are able to characterize and classify a broad variety of random motions,
from classical diffusion and  active run-and-tumble motions to anomalous versions of these, related 
to fractional equations, including also stochastic resetting and trapping (immobilization) mechanisms.
For each type of motion investigated we report the constitutive Fokker-Planck like equation of the process, 
the expressions of the solutions PDFs, the MSDs and their asymptotic behaviors, determining the
anomalous properties of the motion.
The main results are summarized in the Table II. 
We also provide a brief overview of the various types of motion discussed in terms of time-changed processes and subordinators.\\
Some possible developments and perspectives of the present work are listed below.

We have investigated the most simple forms of the function $\g(s)$.
It would be of interest to consider more complex expressions of it
satisfying the condition $\sqrt{\g} \in \mathcal{BF}$ \cite{schilling-bern}
and analyze the possible connection to realistic physical processes.

Our analysis is based on solutions of the Fokker-Planck equations of the form (\ref{Pks}),
which allows us to write the PDF in the $(x,s)$ domain as (\ref{Pxs}) and 
to argue from that the possible allowed forms of the function $\g(s)$.
However, not all random motions are described by the simple solution (\ref{Pks}).
Consider, for example, the alternate motions with different behaviors in the two phases \cite{angelani2013averaged},
the {\it soft} version of the stochastic resetting in the anomalous case \cite{KGN_2019}
or the presence of a generic external potential.
It would be interesting to investigate more general forms of the PDF 
and  extend the analysis to embrace a wider spectrum of random motions.

In this work we have considered equations with spatial operators  $\partial_x^2$,
resulting in  a Lorentzian shape of the solution in the Fourier domain as a function of $k$. 
One could ask what happen in the more general case of different shapes, for example involving higher powers
of $k$ or non-integer powers $|k|^\gamma$. The latter are usually obtained by introducing 
non-local spatial operators into the Fokker-Planck equation, resulting in space-fractional derivatives 
\cite{klafter2011first,kilbas2006theory,OrsZhao2003}.

A further possible extension of this analysis could be the investigation of the $d$-dimensional 
case.  
Simply extending the Fokker-Planck like equation (\ref{FPeq})
to higher spatial dimensions means the presence of a Laplacian, resulting again
in a Lorentzian shape of the Fourier-Laplace transformed PDF (\ref{Pks}) in terms of the modulus $|{\bf k}|$,
but now the Laplace transformed PDF does not take the simple form  (\ref{Pxs}) and we cannot repeat the 
arguments of the present investigation.
Moreover, 
we note that, in some cases, by simply considering the $d$-dimensional version of equation 
(\ref{FPeq}) with the Laplacian operator
is not equivalent to consider random walk models in $d$-dimensional space. 
It would be worthwhile to investigate the extension of the present framework and its underlying arguments to higher spatial dimensions \cite{martens2012probability,Ang_JSP2024,JM_Entropy_2021}.

\acknowledgments
I gratefully acknowledge Roberto Garra for his insightful suggestions and the many stimulating discussions we have had on these topics over the years. 
I also thank Alessandro De Gregorio, Matteo Paoluzzi and Andrea Puglisi
for their helpful comments on the manuscript and for useful discussions.

\begin{sidewaystable}
\label{Tab2}
\centering
\setlength{\tabcolsep}{5pt}
\renewcommand{\arraystretch}{1.5}
 \begin{tabular}{||l| l  r | c| c| c||} 
 \hline 
 \rule{0pt}{20pt} 
 {Type of motion} & \multicolumn{2}{|c|}{$\g(s)$} & Kinetic equation & $\MSD$ &$\MSD$  \\ [-2ex] 
   &  &  & &${\scriptstyle (t\to 0)}$ & ${\scriptstyle (t\to \infty)}$\\ [1ex] 
 \hline 
 \hline
 Diffusion (D) & $s$ &  & $(\partial_t-v^2\partial_x^2)P=0$ &   $t$ & $t$ \\ 
 Ballistic  & $s^2$ &  & $(\partial_t^2-v^2\partial_x^2)P=0$  & $t^2$ & $t^2$ \\ 
 D with resetting/trapping & $s+r$ & & $(\partial_t -v^2\partial_x^2+r)P=r\delta(x)$ &
$t$ & const.  \\
 Anomalous Diffusion (AD) & $s^\mu$ & \begin{tabular}{r} ${\scriptstyle 0<\mu<2}$\end{tabular} & 
 $(\partial_t^\mu -v^2\partial_x^2)P=0$ & 
  $t^\mu$ & $t^\mu$ \\
 AD with resetting & $(s+r)^\mu$ & \begin{tabular}{r} ${\scriptstyle 0<\mu<2}$ \end{tabular}& 
  $\left( \partial_t^\mu  \!-\!v^2 \partial_x^2 \right) e^{rt} P\!=\!\delta(x) J^\mu e^{rt}$
  & 
  $t^\mu$ & const. \\
 AD with trapping & $s^\mu+r$ & \begin{tabular}{r} ${\scriptstyle 0<\mu<1}$\end{tabular} & 
 $(\partial_t^\mu -v^2\partial_x^2+r)P=r\delta(x)$  & 
 $t^\mu$ & const. \\
 Run-and-tumble (RT) & $s (s+\alpha)$ & & 
 $(\partial_t^2 +\alpha \partial_t-v^2\partial_x^2)P=0$ & 
  $t^2$ & $t$ \\
 RT with resetting/trapping & $(s+r)(s+r+\alpha)$ & & 
 $\left(\partial_t^2\!+\!(\alpha\!+\!2r) \partial_t\!-\!v^2\partial_x^2 
 \!+\!r (\alpha\!+\!r)\right)P\!=\!r(\alpha\!+\!r)\delta(x)$ 
 & 
 $t^2$ & const. \\ 
 Anomalous RT (ART)  & $s^\mu (s^\nu +\alpha)$ & 
 \begin{tabular}{r} 
 ${\scriptstyle  0<\mu<2}$\\ [-2.5ex]
 ${\scriptstyle 0<\nu<\min(1,2-\mu)}$  
 \end{tabular}
& $(\partial_t^{\mu+\nu}+\alpha \partial_t^\mu-v^2\partial_x^2)P=0$ & 
$t^{\mu+\nu}$ & $t^\mu$ \\
 ART with resetting & $(s+r)^\mu ((s+r)^\nu+\alpha)$ & 
  \begin{tabular}{r} 
 ${\scriptstyle  0<\mu<2}$\\ [-2.5ex]
 ${\scriptstyle 0<\nu<\min(1,2-\mu)}$  
 \end{tabular}
 & 
 $\left( \partial_t^{\mu+\nu} \!+\! \alpha \partial_t^\mu \!-\! v^2 \partial_x^2\right)e^{rt} P\!=\!
 \delta(x) \left( J^{\mu+\nu}\!+\!\alpha J^\mu \right) e^{rt} $ 
 & $t^{\mu+\nu}$ & const. \\ 
 ART with trapping & $(s^\mu+r)(s^\nu+r+\alpha)$ & 
  \begin{tabular}{r} 
 ${\scriptstyle 0<\mu<1}$\\ [-2.5ex]
 ${\scriptstyle 0<\nu<1}$  
 \end{tabular}
 & 
 $\left( \partial_t^{\mu+\nu}\!+\!(\alpha\!+\!r) \partial_t^\mu +r \partial_t^\nu\!-\!v^2 \partial_x^2 \!+\! r (\alpha\!+\!r)
 \right)  P\!=\!r(\alpha\!+\!r)\delta(x)$ 
 & $t^{\mu+\nu}$ & const. \\ 
\hline
\end{tabular}
\caption{Different random motions  with the corresponding functions $\g(s)$ entering in the expression
 of $P(k,s)$ (\ref{Pks}) and $P(x,s)$ (\ref{Pxs}), and the Fokker-Planck like equations for the probability density function. The operator $J^\mu$ appearing in the equations of anomalous motions with resetting is defined in (\ref{Jop}). We also report the asymptotic behaviors of the MSD for the various models. }
\end{sidewaystable}

\clearpage


\appendix

\section{Some mathematical background}
\label{App}
In this appendix we concisely provide some useful basic definitions and properties of special functions, 
Fourier and Laplace transforms, complete monotone and Bernstein functions and fractional calculus.

\subsection{Special Functions}

\noindent
The {\it gamma function} $\Gamma(z)$ is
\begin{equation}
\label{GammaFun}
    \Gamma(z) = \int_0^\infty dt \ t^{z-1} \ e^{-t} , \qquad {\text Re} z>0 .
\end{equation}
The {\it incomplete gamma function} $\gamma(z,x)$ is 
\begin{equation}
\label{IGammaFun}
    \gamma(z,x) = \int_0^x dt \ t^{z-1} \ e^{-t} , \qquad {\text Re} z>0 .
\end{equation}
The {\it Mittag-Leffler function} $E_\mu$ is defined as \cite{gorenflo2020mittag}
\begin{equation}
E_\mu(z) = \sum_{k=0}^\infty \frac{z^k}{\Gamma(\mu k+1)}   ,
\label{ML1}
\end{equation}
with $\mu \in {\mathbb C}$, Re $\mu>0$.
The Mittag-Leffler function generalizes  the
exponential function, obtained for $\mu=1$,
$E_1(z)=\exp(z)$.\\
The {\it two-parameters Mittag-Leffler function} is a generalization of  (\ref{ML1}) as
\begin{equation}
E_{\mu,\nu} (z) = \sum_{k=0}^\infty \frac{z^k}{\Gamma(\mu k+\nu)}  .
\label{ML2}
\end{equation}
It reduces to (\ref{ML1}) for $\nu=1$, i.e., $E_{\mu,1}(z)=E_\mu(z)$.\\
The {\it Wright function} $W_{\mu,\nu}$ is defined  by
\begin{equation}
W_{\mu,\nu} (z) = \sum_{k=0}^\infty \frac{z^k}{k!\Gamma(\mu k+\nu)}  ,
\label{WF}
\end{equation}
where $\mu>-1$ and $\nu>0$.\\
The {\it Mainardi function} (or $M$-Wright function) $M_{\mu}$  is defined  by
\begin{equation}
M_{\mu} (z) = \sum_{k=0}^\infty \frac{(-1)^k z^k}{k!\Gamma(1-\mu-\mu k)}  ,
\label{MF}
\end{equation}
where $0<\mu<1$. It is related to the Wright function through
$M_\mu(z) = W_{-\mu,1-\mu}(-z)$.\\
The $\mathbb{M}$-Wright function $\mathbb{M}_{\mu}(x,t)$ is defined  by
\begin{equation}
\mathbb{M}_{\mu} (x,t) = t^{-\mu} \ M_\mu \left( x t^{-\mu} \right) ,
\label{MMF}
\end{equation}
with $0<\mu<1$ and $x,t \in \mathbb{R}^+$, which defines a PDF in $x$ evolving in $t$ with self-similarity exponent $H=\mu$ \cite{Mainardi_Rev_2010}.\\
The {\it modified Bessel functions} $I_\nu(z)$ of order $\nu \in {\mathbb R}$  are defined as \cite{gradshteyn2014table}
\begin{equation}
    \label{MBF}
    I_\nu(z) = \sum_{k=0}^{\infty} \frac{1}{k! \Gamma(\nu+k+1)} \left( \frac{z}{2}\right)^{2k+\nu} .
\end{equation}

\subsection{Fourier and Laplace transforms}
\label{Sub_FLT}

The {\it Fourier transform} of a function $f(x):{\mathbb R} \to {\mathbb R} $ is defined as 
\begin{equation}
\label{AFT}
    {\hat f} (k) =
    {\mathcal F}[f(x)](k) = 
    \int_{-\infty}^{+\infty} dx \ e^{i k x} \ f(x) .
\end{equation}
The inverse Fourier transform is
\begin{equation}
\label{AIFT}
    f (x) =
        {\mathcal F}^{-1}[{\hat f}(k)](x) = \frac{1}{2\pi}
    \int_{-\infty}^{+\infty} dk \ e^{- i k x} \ 
    {\hat f}(k) .
\end{equation}
Some basic properties of the Fourier transform pairs $f(x) \div {\hat f}(k)$ are
\begin{gather}
\label{FT_lin}
a f(x) + b g(x) \div a {\hat f}(k) + b {\hat g}(k) ,\\
\label{FT_der}
f^{(n)}(x)  \div (-ik)^n {\hat f} (k) , \\
\label{FT_conv}
( f * g )(x) \div {\hat f}(s) \cdot {\hat g}(s) , 
\end{gather}
where in (\ref{FT_lin}) $a$ and $b$ are arbitrary constants, in (\ref{FT_conv}) 
$$(f*g)(x)=\int_{-\infty}^{+\infty} d\xi f(x-\xi) g(\xi)$$
is the convolution integral and in (\ref{FT_der})
$f^{(n)}$ denotes the $n$-th derivative of $f$ and we have assumed that 
$\lim_{|x|\to \infty } f^{(m)}(x) = 0$ for $m=0,1,\dots,n-1$.\\
Useful transform pairs  used in this work are
\begin{align}
\label{FT1}
\delta(x) &\div 1 \\
\label{FT2}
e^{-a|x|} &\div \frac{2a}{a^2+k^2}  
 \hspace{1.8cm} {\scriptstyle (a>0) } \\
\label{FT3}
e^{-a^2 x^2} &\div \frac{\sqrt{\pi}}{a} \ e^{-k^2/4a^2}  
 \hspace{1.8cm} {\scriptstyle (a>0) } 
 \end{align}

\noindent
The {\it Laplace transform} of a function $f(t): [0,\infty) \to {\mathbb R}$ is defined as
\begin{equation}
\label{ALT}
    {\tilde f} (s) =
    {\mathcal L}[f(t)](s) = 
    \int_0^\infty dt \ e^{-st} \ f(t) ,
\end{equation}
where Re $s >0$.
Some useful properties of the Laplace transform pairs $f(t) \div {\tilde f}(s)$ are as follows
\begin{gather}
\label{LT_lin}
a f(t) + b g(t) \div a {\tilde f}(s) + b {\tilde g}(s) ,\\
\label{LT_trasl}
e^{-at} f(t)   \div {\tilde f}(s+a) , \\
\label{LT_der}
f^{(n)}(t)  \div s^n {\tilde f} (s) - s^{n-1} f(0) - s^{n-2} f^{(1)}(0) - \dots - f^{(n-1)}(0) , \\
t^n f(t) \div (-1)^n {\tilde f}^{(n)}(s) ,\\
\label{LT_int}
\int_0^t d\tau\ f(\tau) \div s^{-1} {\tilde f}(s) ,\\
\label{LT_conv}
( f * g )(t) \div {\tilde f}(s) \cdot {\tilde g}(s) , 
\end{gather}
where in (\ref{LT_lin}) $a$ and $b$ are arbitrary constants, 
in (\ref{LT_der}) $f^{(n)}$ is the $n$-th derivative of $f$ and we have assumed that
$\lim_{t\to \infty} f(t) e^{-st} = 0$ and 
in (\ref{LT_conv})
$$(f*g)(t)=\int_0^t d\tau f(t-\tau) g(\tau)$$
is the convolution integral.\\
Some  Laplace transform pairs 
used in this work are:
\begin{align}
\label{LT_delta}
\delta(t) &\div 1 \\
\label{LT_delta2}
 \delta(t-\tau) &\div e^{-\tau s} \\
 \label{LT_1sus}
1 &\div \frac{1}{s}\\
\label{LT_tn1}
t^n &\div \frac{n!}{s^{n+1}} \\
\label{LT_tn2}
 t^n e^{-at} &\div \frac{n!}{(s+a)^{1+n}} \\
\label{LT_D}
 t^{-1/2} e^{-a/t} &\div \sqrt{\pi} s^{-1/2}  e^{-2\sqrt{as}}
  \hspace{0.8cm} {\scriptstyle (a \geqslant 0)} \\
\label{LT_pow}
 t^\mu &\div \frac{\Gamma(1+\mu)}{s^{1+\mu}} 
 \hspace{1.8cm} {\scriptstyle (\mu>-1) } \\
  \label{LT_sh}
 \sinh{at} &\div \frac{a}{s^2-a^2} \\
\label{LT_ch}
  \cosh{at} &\div \frac{s}{s^2-a^2} \\
 \label{LT_E1}
 E_\mu(at^\mu) &\div \frac{s^{\mu-1}}{s^\mu-a} 
 \hspace{2.15cm} {\scriptstyle (\mu>0) } \\
 \label{LT_E2}
 t^{\nu-1} E_{\mu,\nu}(a t^\mu) &\div \frac{s^{\mu-\nu}}{s^\mu-a} 
 \hspace{2.15cm} {\scriptstyle (\mu,\nu>0) }
\end{align}
The reader can refer to Ref.s \cite{gradshteyn2014table} and \cite{gorenflo2020mittag} for
more details and examples.\\

\noindent
There is a relation between the asymptotic behavior of a function and that of its Laplace transform, given by Tauberian theorems 
\cite{feller1971,klafter2011first,Bressloff_book}.\\
\noindent
{\it Tauberian theorem. }
If  $f(t)\geqslant0$ is an ultimately monotonic function, $0\leqslant \nu < \infty$
and $L(t)$ a slowly varying function for $t\to \infty$, each of the following  relations implies the other
\begin{align}
\label{Tuab1}
    f(t) &\simeq L(t) \ t^{\nu-1}, \qquad t \to \infty ,\\
    {\tilde f}(s) &\simeq \Gamma(\nu) L(1/s)\  s^{-\nu} , \qquad s \to 0 .
\label{Taub2}
\end{align}
The same is true considering the opposite limit $t\to 0$ and $s \to \infty$, with $L(t)$ a slowly varying function
at zero \cite{feller1971,klafter2011first}.

\subsection{Completely monotone and Bernstein functions}

{\it Completely monotone functions} ($\mathcal{CM}$).
A function $f(s):(0,\infty) \to {\mathbb R}$ is a 
{completely monotone function} 
if
$f$ is of class $C^\infty$ and, for all $s>0$ and $n\in \mathbb{N}\cup \{0\}$
$$
(-1)^n f^{(n)}(s) \geqslant 0 \qquad \qquad (\mathcal{CM}) ,
$$
where  $f^{(n)}(s)$ denotes the derivative of order $n$.
Examples of $\mathcal{CM}$  are: $1$, $1/s$, $(s+a)^{-1}$ with $a>0$, $e^{-as}$ wiht $a>0$, 
$s^{-\mu}$ with $\mu>0$.\\
{\it Bernstein's theorem}. A function $f(s):(0,\infty) \to {\mathbb R}$ is $\mathcal{CM}$ if and only if it is the Laplace
transform of a non-negative measure $\mu(dt) = m(t) dt$ 
\begin{equation}
    f\in {\mathcal{CM}} \Longleftrightarrow f(s) = \int_0^\infty e^{-st} \mu(dt) .
\end{equation}

{\it Stieltjes functions} (${\mathcal S}$). 
{Stieltjes functions} 
are a subclass of completely monotone functions 
(${\mathcal S}\subset \mathcal{CM}$) and 
are defined by the following representation
\begin{equation}
\label{SF}
f\in {\mathcal{S}} \Longleftrightarrow  f(s) = \frac{a}{s} + b + \int_0^\infty \frac{1}{s+t}\ \mu(dt) \ ,
\end{equation}
with $a,b\geqslant0$, $\mu(dt)=m(t)dt$ is a measure on $(0,\infty)$ and the integral converge.
${\mathcal S}$ consists of all $\mathcal{CM}$ functions which have a representation measure that
is $\mathcal{CM}$.
Examples of $\mathcal{S}$  are: $1$, $1/s$, $(1+a)/(s+a)$ with $a>0$,  
$s^{\mu-1}$ with $0<\mu<1$, $s^{-1} \log(1+s)$.\\

{\it Bernstein function} ($\mathcal{BF}$).
A function $f(s):(0,\infty) \to {\mathbb R}$ is a 
{Bernstein function} 
if $f$ is of class $C^\infty$, $f(s)\geqslant 0$ for all $s>0$
and, for all $s>0$ and $n\in \mathbb{N}$
$$
(-1)^{n-1} f^{(n)}(s) \geqslant 0 \qquad \qquad (\mathcal{BF}) .
$$
A function $f$ is $\mathcal{BF}$ if and only if 
admits the L\`evy-Khintchine representation
\begin{equation}
\label{LBF}
f\in {\mathcal{BF}} \Longleftrightarrow  f(s) = a + b s + \int_0^\infty (1-e^{-st}) \ \mu(dt) \ ,
\end{equation}
with $a,b\geqslant0$, $\mu(dt)=m(t)dt$ is a measure on $(0,\infty)$ and the integral converge.
The triplet $(a,b,\mu)$ is called the L\`evy measure of $f$.\\
Examples of $\mathcal{BF}$  are: $1$, $s$, $s/(1+s)$, 
$(1+1/a)(1-e^{-sa})$ with $0\leqslant a<\infty$, $\log(1+s)$,  
$s^{\mu}$ with $0<\mu<1$.\\

{\it Complete Bernstein functions} (${\mathcal{CBF}}$). 
A Bernstein function $f$ is said to be {complete Bernstein function} 
if its L\`evy measure in (\ref{LBF}) has a complete monotone density 
$m\in \mathcal{CM}$. We have that $\mathcal{CBF}\subset \mathcal{BF}$.
Every function $f \in \mathcal{CBF}$ can be uniquely represented in the form 
(Stieltjes representation)
\begin{equation}
\label{CBF}
f\in {\mathcal{CBF}} \Longleftrightarrow  f(s) = a + b s+ \int_0^\infty \frac{s}{s+t}\ \mu(dt) \ ,
\end{equation}
with $a,b\geqslant0$, $\mu(dt)=m(t)dt$ is a measure on $(0,\infty)$ and the integral converge.\\
Examples of $\mathcal{CBF}$  are: $1$, $s$, $s(1+a)/(s+a)$ with $0\leqslant a<\infty$, 
$\log(1+s)$, $s^{\mu}$ with $0<\mu<1$.\\

Some useful properties are as follows.
\begin{itemize}
\item[({\it i})] If $f,g\in {\mathcal{I}}$ then $af+bg \in {\mathcal I}$ for all $a,b\geqslant 0$, 
where ${\mathcal I} \in \{\mathcal{CM, S, BF, CBF}\}$
\item[({\it ii})] If $f,g\in {\mathcal{CM}}$ then $f \cdot g \in {\mathcal{CM}}$.
\item[({\it iii})] If $f,g\in {\mathcal{BF}}$ then $g \circ f \in {\mathcal{BF}}$.
\item[({\it iv})] $f\in {\mathcal{BF}}$ if and only if $g \circ f \in {\mathcal{CM}}$ for every $g\in {\mathcal{CM}}$.
\item[({\it v})] If $f\in {\mathcal{BF}}$ then $f/s \in {\mathcal{CM}}$.
\item[({\it vi})] $f\in {\mathcal{BF}}$ if and only if $e^{-af} \in {\mathcal{CM}}$ for every $a>0$.
\item[({\it vii})] If $f,g\in {\mathcal{BF}}$ then $f(s^\alpha) \cdot g(s^{\beta}) \in {\mathcal{BF}}$ 
with $\alpha,\beta \in (0,1)$ and $\alpha+\beta\leqslant1$.
\item[({\it viii})] If $f,g\in {\mathcal{CBF}}$ then $g \circ f \in {\mathcal{CBF}}$.
\item[({\it ix})] If $f,g\in {\mathcal{CBF}}$ then $f^\alpha \cdot g^{1-\alpha} \in {\mathcal{CBF}}$ with $\alpha\in(0,1)$.
\item[({\it x})] If $f,g\in {\mathcal{S}}$ then $f^\alpha \cdot g^{1-\alpha} \in {\mathcal{S}}$ with $\alpha\in(0,1)$.
\item[({\it xi})] If $f,g\in {\mathcal{CBF}}$ then $(f^\alpha(s)+ g^\alpha(s))^{1/\alpha} \in {\mathcal{CBF}}$ 
with $\alpha\in [-1,1] \backslash \{0\}$.
\item[({\it xii})] If $f,g\in {\mathcal{CBF}}$ then $(f(s^\alpha)+ g(s^\alpha))^{1/\alpha} \in {\mathcal{CBF}}$ 
with $\alpha\in [-1,1] \backslash \{0\}$ (then also $f(s^\alpha)^{1/\alpha} \in {\mathcal{CBF}}$).
\item[({\it xiii})] If $f,g\in {\mathcal{CBF}}$ then $f(s^\alpha) \cdot g(s^{1-\alpha}) \in {\mathcal{CBF}}$ 
with $\alpha\in [0,1]$.
\item[({\it xiv})] $f\in {\mathcal{CBF}}$ if and only if $f/s \in {\mathcal{S}}$.
\item[({\it xv})] $f\in {\mathcal{CBF}}$ (with $f \not\equiv 0$) if and only if $1/f \in {\mathcal{S}}$.
\end{itemize}
We refer to  Ref.\cite{schilling-bern} for a complete discussion of this topic.

\subsection{Fractional Calculus}
We define the {\it Riemann-Liouville fractional integral} 
(also known in the literature as Abel-Riemann fractional integral) 
of order $\mu>0$ as \cite{gorenflo2020mittag}
\begin{equation}
\label{FracIntegral}
    I^{\mu} f(t) = \frac{1}{\Gamma(\mu)} \int_0^t d\tau \ (t-\tau)^{\mu-1} f(\tau) ,
\end{equation}
which generalizes the $n$-fold primitive of a function $f$ to a non integer order.
We put $I^0 = {\mathbb I}$ (identity operator). 
The following property holds
\begin{equation}
    I^\mu I^\nu = I^{\mu+\nu} \qquad (\mu,\nu\geqslant0) .
\end{equation}
The {\it Riemann-Liouville fractional derivative} is defined as
\begin{equation}
    D^\mu f(t) =
        \frac{1}{\Gamma(n-\mu)} \frac{d^n}{dt^n} \int_0^t d\tau \ (t-\tau)^{n-1-\mu} f(\tau) , \quad n-1<\mu<n ,
\end{equation}
where $n\in {\mathbb N}$.
For integer order $\mu=n$ we have 
\begin{equation}
   D^n f(t) = \frac{d^n}{dt^n} f(t), \qquad n\in {\mathbb N} . 
\end{equation}
We put $D^0={\mathbb I}$. 
The operator $D^\mu$ is the left-inverse of $I^\mu$, i.e.,
\begin{equation}
\label{FD_identity}
    D^\mu I^\mu = {\mathbb I} .
\end{equation}
It is useful to introduce another fractional operator, the {\it Caputo fractional derivative}
${^CD}^\mu$
(in the main text we indicate the Caputo fractional partial derivative with $\partial_t^\mu$)
, as
\begin{equation}
\label{CapDer}
    {^CD}^{\mu} f(t) =
        \frac{1}{\Gamma(n-\mu)} \int_0^t d\tau \ (t-\tau)^{n-1-\mu} f^{(n)}(\tau) , \quad \quad n-1<\mu<n ,
\end{equation}
and, for integer $\mu=n$
\begin{equation}
   {^CD}^n f(t) =\frac{d^n}{dt^n} f(t), \qquad n\in {\mathbb N} . 
\end{equation}
For $n=1,2$ the Caputo derivatives read 
\begin{align}
\label{Capdev1}
{^CD}^{\mu} f(t) &= \frac{1}{\Gamma(1-\mu)} \int_0^t d\tau \ (t-\tau)^{-\mu} f^{(1)}(\tau) , \quad \quad 0<\mu<1 , \\
\label{Capdev2}
{^CD}^{\mu} f(t) &= \frac{1}{\Gamma(2-\mu)} \int_0^t d\tau \ (t-\tau)^{1-\mu} f^{(2)}(\tau) , \quad \quad 1<\mu<2 .
\end{align}
We have the following relation between the Riemann-Liouville fractional integral and the Caputo fractional derivative
\begin{equation}
\label{CIrel}
 {^CD}^{\mu} = I^{n-\mu} D^{n}, \qquad n-1<\mu<n .
\end{equation}
The Caputo and Riemann-Liouville fractional derivatives differ by boundary terms
$$
D^\mu f(t) = {^CD}^{\mu} f(t) + \sum_{k=0}^{n-1} \frac{t^{k-\mu}}{\Gamma(1+k-\mu)} f^{(k)}(0) ,
$$
where $n-1<\mu<n$.\\
We have the following properties of the Laplace transforms, which generalize (\ref{LT_der}) and (\ref{LT_int})
to the fractional case
\begin{align}
\label{CFD_p1}
    {^CD}^{\mu} f(t) &\div s^\mu {\tilde f}(s) - \sum_{k=0}^{n-1} s^{\mu-1-k} f^{(k)}(0) , \qquad n-1<\mu<n , \\
    I^{\mu} f(t) &\div s^{-\mu} {\tilde f}(s) ,  \qquad \mu>0 .
\label{CFD_p2}
\end{align}
The reader can consult Ref.\cite{gorenflo2020mittag} (Appendix E) for a short introduction to fractional calculus and Ref.\cite{kilbas2006theory} for an in-depth discussion of the topic.

  \input{ref.bbl}


\end{document}

%% file: ref.bbl
%